\documentclass[12pt]{article}

\usepackage{amsmath}
\usepackage{amsfonts}
\usepackage{amssymb}

\usepackage{graphicx}
\usepackage{epsfig}
\usepackage{epsf}

\allowdisplaybreaks

\begin{document}

\author{A. Nehme}
  
\title{Isospin Breaking in $K_{\ell 4}$ Decays of the \\
       Neutral Kaon}
\maketitle

\begin{center}
Centre de Physique Th\'eorique \\
Luminy, case 907 \\
F-13288 Marseille Cedex 09
\end{center}

\begin{center}
\texttt{nehme@cpt.univ-mrs.fr} 
\end{center}
 
\begin{abstract}
In the presence of photons, the neutral $K_{\ell 4}$ decay,
$K^0\rightarrow\pi^0\pi^-\ell^+\nu_\ell$, can be parameterized in
terms of three vectorial, one anomalous, and one tensorial form
factors. We present here analytic expressions of two vectorial
form factors, $f$ and $g$, calculated at one-loop level in the
framework of chiral perturbation theory based on the effective
Lagrangian including mesons, photons, and leptons. These expressions 
may then be used to disentangle the Isospin breaking part from the measured form factors and hence improve the accuracy in the determination of $\pi\pi$ scattering parameters from $K_{\ell 4}$ experiments.
\end{abstract}

\textbf{keywords:} Electromagnetic Corrections, Kaon Semileptonic Decay, Form Factors, Chiral Perturbation Theory.

\pagebreak 

\tableofcontents

\section{Introduction}
\label{introduction} 

The fundamental state for Quantum Chromodynamics (QCD) remains a poorly
understood medium due to its complicated non perturbative
structure. It is widely believed that one of the main features of
the vacuum structure is color confinement. In the chiral limit,
defined by the vanishing of the light quark masses, confinement
leads to the spontaneous breaking of chiral symmetry provided that
the number of light flavors, $N_f$, is at least three~\cite{Coleman:1980mx}. Assuming
further that the QCD vacuum angle vanishes, then, the fundamental
state is invariant under the vector (residual) symmetry~\cite{Vafa:1984tf}. As stated
by Goldstone's theorem~\cite{Goldstone:1962es}, $N_f^2-1$ massless pseudoscalars,
$\phi^a$, are created from the vacuum and coupled to the axial
currents, $A_{\mu}^a$, through 
\begin{equation}
\langle 0|A_{\mu}^a(0)|\phi^b(p)\rangle\,=\,i\delta^{ab}F_0p_{\mu}\,. 
\end{equation}
The pseudoscalars form a non linear realization of chiral symmetry~\cite{Coleman:1969sm} which becomes linear when restricted to vector transformations~\cite{Callan:1969sn}. 
Furthermore, their dynamics is dictated by the present symmetry
breaking pattern which constrains them to interact via derivative
couplings~\cite{Burgess:1998ku}. Relaxing the hypothesis of vanishing quark masses in
favor of the assumption that quarks are massive but light, the
pseudoscalars acquire dynamical masses by means of the vacuum
alignment mechanism~\cite{Dashen:1971et, Weinberg:1976gm}. 
This constitutes the basis for a systematic
low-energy expansion in powers of $p/\Lambda$ and $m_q/\Lambda$
where $q$ stands for a light quark ($u$, $d$, and probably $s$), and
$\Lambda$ is the mass threshold for producing particles heavier
than the pseudoscalars. Taking $N_f=3$ and identifying the
pseudoscalars with the eight light mesons of the hadronic
spectrum, the preceding (chiral) expansion is nothing else than
the effective field theory of QCD and is called Chiral
Perturbation Theory (ChPT)~\cite{Langacker:1973hh, Pagels:1975se}. 
All the features of the underlying theory,
and particularly its symmetry breaking pattern, are coded in the
coefficients of the chiral expansion which are the parameters of
the effective Lagrangian. Thereby, they are called order
parameters of the chiral symmetry since they point out its
spontaneous breaking. Among the infinitely many order parameters
the coupling $F_0$ and the quark condensate, $\langle{\overline
q}q\rangle$, are of particular interest. While the non vanishing
of the former is necessary and sufficient for the spontaneous
breaking of chiral symmetry, a non zero value for the latter
breaks chiral symmetry spontaneously but the converse is not
necessarily true~\cite{Stern:1998dy}. Chiral symmetry constraints are unable to
predict the values of order parameters. In return, large $N_c$ counting rules
plead in favor of a sizable quark condensate. In order to confirm
or reject this argument, a phenomenological determination of order
parameters is primordial. With respect to the coupling $F_0$, it can be extracted
from the pion decay constant, 
\begin{equation}
F_{\pi}\,=\,F_0\left[ 1+{\cal O}(m_q)\right ]\,. 
\end{equation} 
Concerning the quark condensate, it enters the
expression of all observables through meson masses~\cite{Gell-Mann:1968rz}, 
\begin{equation}
\label{eq:mass_expansion} 
F_{\pi}^2M_{\pi}^2\,=\,-(m_u+m_d)\langle{\overline q}q\rangle
+{\cal O}(m_q^2)\,. 
\end{equation} 
Therefore, one cannot determine the size
of this order parameter without a preliminary assumption on the
chiral order of quark masses. The large $N_c$ prediction means
that the pion mass is dominated by the contribution from the quark
condensate. Consequently, the first term in the right-hand-side of
(\ref{eq:mass_expansion}) is leading order, that is, ${\cal O}(m_q)={\cal
O}(p^2)$. It follows that the ratio 
\begin{equation} 
\label{eq:flavor_ratio}
X\,\doteq\,-\frac{(m_u+m_d)\langle{\overline q}q\rangle}{F_{\pi}^2M_{\pi}^2}\,,
\end{equation} 
is proving a suitable parameter to test the present power
counting scheme. Important deviations of $X$ from $1$ should
incite on a careful reexamination of the validity of large $N_c$
arguments in the vacuum sector. It is interesting then to perform
accurate measurements of observables that are sensitive to
variations of $X$. This is the case of low-energy $\pi\pi$
scattering. In fact, the latter is solely described in the
threshold region in terms of the $S$-wave scattering lengths, 
\begin{eqnarray}
a_0^0 
&=& \frac{M_{\pi}^2}{96\pi F_{\pi}^2}\left[ 5\alpha +16\beta +\mathcal{O}(m_q)\right] \,,
\\
a_0^2 
&=& \frac{M_{\pi}^2}{48\pi F_{\pi}^2}\left[ \alpha -4\beta +\mathcal{O}(m_q)\right] \,,
\end{eqnarray} 
where,
\begin{equation}
\alpha\,=\,4-3X+\mathcal{O}(m_q)\,, \quad \beta\,=\,1+\mathcal{O}(m_q)\,.
\end{equation}
For instance, the isoscalar scattering length,
$a_0^0$, varies by a factor $2$ when $X$ varies from $1$ to zero.
This sensitivity of $\pi\pi$ scattering to the size of the quark
condensate is not only a feature of the leading order but, on the
contrary, it persists at higher orders allowing $\pi\pi$
scattering to be the golden process for testing the mechanism of
quark condensation~\cite{Knecht:1995tr}. 

The $\pi\pi$ scattering is experimentally
accessible in $K_{\ell 4}$ and pionium decays. In fact, the partial wave
expansion of $K_{\ell 4}$ form factors displays the $\pi\pi$ phase
shifts as stated by the Watson final state interaction theorem~\cite{Watson:1952ji, Watson:1954uc}. The data on phase shifts can then be translated into a
model-independent determination of scattering lengths by means of
Roy equations~\cite{Ananthanarayan:2000ht}. Concerning the pionium characteristics such as its
lifetime, $\tau$, and $2S-2P$ strong energy level shift, $\Delta E_s$, they give direct access to $\pi\pi$ scattering lengths via~\cite{Efimov:1986},
\begin{equation} 
\label{eq:pionium} 
\tau\,\propto\,(a_0^0-a_0^2)^2\,, \quad \Delta
E_s\,\propto\,2a_0^0+a_0^2\,. 
\end{equation} 
Once the results from the
presently running DIRAC experiment are available, $a_0^0$ and
$a_0^2$ should be determined with $5$ to $10\%$ accuracy~\cite{Adeva:2003up}. On the
other hand, the charged $K_{\mathrm{e}4}$ decay has been measured
by the E$865$ experiment~\cite{Pislak:2001bf, Pislak:2003sv} and the outgoing data have been analyzed in~\cite{Colangelo:2001sp} and~\cite{Descotes-Genon:2001tn},
independently. Before recalling the conclusions of both
references, let us stress that the obtained value for $a_0^0$ is
$7\%$ accuracy and is compatible with the prediction of the
standard power counting which rests on large $N_c$ grounds. The
analysis of Ref.~\cite{Colangelo:2001sp} relies on chiral symmetry
constraints which correlate the two scattering lengths. It yields
a value for $a_0^2$ consistent with the standard counting
prediction. Furthermore, if one combines the extracted values for
the scattering lengths, the ratio (\ref{eq:flavor_ratio}) reads then,
$X\sim 0.94$. As for the analysis performed in
Ref.~\cite{Descotes-Genon:2001tn}, it combines the data of the E$865$
experiment with an existing one in the isospin-two channel below
$800$ MeV without using the aforementioned chiral constraints. The
conclusions of this analysis point out a discrepancy at the
$1-\sigma$ level between the extracted value for $a_0^2$ and the
one predicted by the standard power counting. Moreover, the
corresponding value for the ratio (\ref{eq:flavor_ratio}) was found to be,
$X\sim 0.81$. In view of the ``disagreement'' between the results of
the two analysis, and before drawing off any conclusion about the
size of the quark condensate, it is necessary to devote much more
effort at both the experimental and the theoretical levels. In
this direction, new precise measurements of charged and neutral $K_{\mathrm{e}4}$ decay are currently taking data at CERN~\cite{Batley:2000} and FNAL~\cite{Santos:2003}, respectively. This should be accompanied by an improvement of the accuracy in the theoretical prediction for the
scattering lengths by evaluating isospin breaking effects in
$\pi\pi$ scattering as well as in $K_{\ell 4}$ decays. While such
effects in the former case are now under control at leading and
next-to-leading orders~\cite{Maltman:1997nw, Maltman:1997nwE, Meissner:1997fa, Meissner:1997faE, Knecht:1998jw, Knecht:2002gz}, we are interested in evaluating these effects at the same orders for the latter case. The aim of the present work is to evaluate isospin breaking effects in the $K_{\ell 4}$ decay of the neutral kaon. The evaluation of the same effects in the $K_{\ell 4}$ decay of the charged kaon are now under control and will be published soon.

\section{$K_{\ell 4}$ decays in the isospin limit} 
\label{sec:isospin_limit}

The semileptonic $K_{\ell 4}$ decays are given schematically by, 
\begin{equation}
\label{eq:processes} 
K(p)\longrightarrow\pi (p_1)\,\pi
(p_2)\,\ell^+(p_l)\,\nu_{\ell}(p_{\nu})\,, 
\end{equation} 
where the lepton, $\ell$, is either a muon, $\mu$, or an electron, e, and
$\nu$ stands for the corresponding neutrino.

Let us start with a historical survey on $K_{\ell 4}$ decays calculation taking into account two important events (hence, three periods): The advent of Current Algebra and that of Effective Lagrangians.  

\textit{Before Current Algebra}. The study of $K_{\ell 4}$ decays starts in the late fifties~\cite{Oneda:1959} with an attempt to extract some information about the decay rate from comparison of the available volume in phase space and of the matrix elements for $K_{\ell 4}$ and $K_{\ell 3}$ decays. The interest was to test some models of the strong interaction. In this direction, the decay rate was computed on the basis of the universal Fermi interaction and some restrictions was imposed by strong interaction selection rules~\cite{Okun:1959pe}. The study of $K_{\ell 4}$ decays proceeded with the calculation of the energy spectrum for leptons~\cite{Chadan:1959on}, the energy spectrum for pions~\cite{Mathur:1959}, and the distribution over the angle between the directions of emission of the pion and the lepton~\cite{Shabalin:1961}. The preceding studies were performed neglecting the interaction of pions in the final state. Deviations from this statement were considered in~\cite{Giocchetti:1962} and~\cite{VanHieu:1963} using dispersion relation techniques. The results obtained in~\cite{Giocchetti:1962} showed that the shape of the momentum distribution is sensitive to the final state interaction of pions while the decay rate is affected slightly. On the other hand, it has been shown in~\cite{VanHieu:1963} that, under reasonable assumptions, the $K_{\ell 4}$ decay mass spectra for the two pions are completely determined by the partial $S$- and $P$-wave amplitudes for $K\overline{K}\rightarrow\pi\pi$ scattering. Hence, experimental data on $K_{\ell 4}$ decays can be used to obtain some definite information on $\pi\pi$ and $\pi K$ interactions. To this end, calculations were made again under the assumption that $K_{\ell 4}$ decays are mediated by pion-pion and pion-kaon resonances in~\cite{Arbuzov:1963, Kacser:1965} or sigma resonance in~\cite{Brown:1964}. Another approach was proposed in~\cite{Shabalin:1963, Shabalin:1963E} pointing out the usefulness of correlations between angular variables describing $K_{\ell 4}$ decays in the determination of the difference between the $S$-wave and $P$-wave $\pi\pi$ scattering phase shifts. This approach was then detailed in~\cite{Cabibbo:1965, Cabibbo:1965E} with a full kinematical description of $K_{\ell 4}$ decays. Finally, a complete review about $K_{\ell 4}$ decay studies during this period can be found in~\cite{Lee:1966jt, Lee:1966hp}.  

\textit{The current Algebra period}. The first Current Algebra calculation of $K_{\ell 4}$ decay form factors was made in~\cite{Callan:1966hu} for pion momenta put equal to zero. Different values were obtained for one form factor depending on which of the two pions in the final state was taken to be soft. The calculation was reconsidered by the author of~\cite{Weinberg:1966, Weinberg:1966E} who explained that the aforementioned difference is due to a missing piece in the previous calculation corresponding to a $K$ pole. The residue of the latter being determined from $\pi K$ scattering and has been discussed in~\cite{Berman:1968ss}. The Current Algebra values for form factors have then been used to predict decay rates~\cite{Berends:1967}. In view of the Current Algebra prediction for the form factors, and with intent to test this prediction experimentaly, it is of considerable interest to determine form factors from experiment in a direct and economical way. It has been shown in~\cite{Pais:1968, Berends:1968} that both form factors and $\pi\pi$ phase shift difference can be directly obtained from the measurement of the intensity spectrum under the assumption that $S$- and $P$-wave dipion states dominate over the higher partial waves in $K_{\ell 4}$ decays. The decay rates calculated from the values of form factors predicted by Current Algebra showed some discrepancy with the experimentally measured ones at that time. In order to account for this discrepancy in the $K_{\mathrm{e}4}$ case, the form factors were calculated again~\cite{Nasrallah:1970ta} by extrapolating decay amplitudes from soft pion limits to the physical point using collinear dispersion relation techniques. The same method was applied later to the $K_{\mu 4}$ case in~\cite{Becherrawy:1976ij}. For a complete review about the use of Current Algebra to study $K_{\ell 4}$ decays, we refer to~\cite{Chounet:1972yy}. 

\textit{The Effective Lagrangian period}. The first application of the effective chiral Lagrangian method to the calculation of $K_{\ell 4}$ form factors was done in~\cite{Ebert:1979rj} using the non-linear realization of chiral symmetry. The calculation of form factors with the help of an effective Lagrangian with linear realization of chiral symmetry was done in~\cite{Shabalin:1989gw} where the experimental values for decay rates has been reproduced thanks to the mixing of scalar $\overline{q}q$ mesons to gluonium. Within the framework of ChPT, the $K_{\ell 4}$ form factors were calculated in~\cite{Bijnens:1990mr, Riggenbach:1991zp, Bijnens:1994ie} to next-to-leading chiral order and in the isospin limit. A fit between the obtained results and the experimental values led to a determination of low-energy constants of the chiral Lagrangian under large-$N_c$ assumption. A model-dependent calculation of form factors was carried out in~\cite{Finkemeier:1993tk} using chiral Lagrangian with hidden local symmetry where vector mesons, $J^P=1^-$, play the role of gauge bosons. The results were in significantly worse agreement with experiment compared with ChPT results and suggested the inclusion of a scalr $J^P=0^+$ resonance in the $\pi\pi$ channel. In the framework of generalized ChPT, where the ratio (\ref{eq:flavor_ratio}) is a free parameter, the $K_{\ell 4}$ form factors were computed to leading order by the authors of~\cite{Knecht:1993eq} who showed that high precision $K_{\mu 4}$ experiments should allow for a direct measurement of the quark mass ratio $m_s/(m_u+m_d)$. In view of the increasing accuracy offered by the one-loop level evaluation of form factors, a reconsideration of the method proposed in~\cite{Pais:1968} to extract $\pi\pi$ phase shifts from $K_{\ell 4}$ decay experiments is indispensable. This was achieved in~\cite{Colangelo:1994qy} with the conclusion that next-to-leading order corrections to the form factors induce a correction to the method not exceeding the $1\%$. The first estimation of two-loop corrections to the $K_{\ell 4}$ form factors was made by the authors of~\cite{Bijnens:1994ie} using dispersion relations. The same subject was also considered by the author of~\cite{Hannah:1995si} with the help of the inverse-amplitude method. He used unitarity and dispersion relations together with the chiral expansion to investigate the effects of two-loop corrections coming from $\pi\pi$ rescattering. The one-loop and partial two-loop results have then been used by the authors of~\cite{Amoros:1999mg} who proposed another method to extract $\pi\pi$ phases from $K_{\ell 4}$ decay data. The method assumes a linear parametrization for the dependence of form factors on kinematical variables and allows a better exploitation of data than the one described in~\cite{Pais:1968, Colangelo:1994qy}. Finally, the full two-loop level calculation of $K_{\ell 4}$ form factors has been done in~\cite{Ametller:1993hg, Amoros:2000mc, Amoros:2000mcE}.              
         
\subsection{Matrix elements}

The decays (\ref{eq:processes}) are described in terms of an invariant
decay amplitude, ${\cal A}$, defined via the matrix element,
\begin{eqnarray}
& & \langle\pi (p_1)\pi (p_2)\ell^+(p_l)\nu_{\ell}(p_{\nu})|K(p)\rangle
\nonumber \\ 
& & \qquad\qquad \doteq\,i\,(2\pi )^4\delta^{(4)}(p_1+p_2+p_l+p_{\nu}-p)\left (-i\,{\cal A}\right )\,,
\end{eqnarray} 
with the on-shell conditions, 
\begin{equation}
p^2\,=\,M^2\,, \; p_1^2\,=\,M_1^2\,, \; p_2^2\,=\,M_2^2\,, \; p_l^2\,=\,m_l^2\,, \; p_{\nu}^2\,=\,0\,. 
\end{equation} 
In the absence of electromagnetism, the decay (\ref{eq:processes}) proceeds
at leading order in Perturbation Theory through the exchange of a
$W^{\pm}$ boson between two leptonic left-handed currents. For
energies very small compared to $M_{W^{\pm}}$, such decays can
then be described by the effective local
Hamiltonian, 
\begin{equation} 
{\cal H}_{\mathrm{eff}}\,=\,\frac{G_F}{\sqrt{2}}\,{\cal
J}_{\mu}^{\dagger} \sum_{\ell}{\overline
\nu}_{\ell}\gamma^{\mu}(1-\gamma^5)\ell +\,\mathrm{hermitian~conjugate}\,. 
\end{equation} 
In the preceding equation, 
\begin{equation}
\label{eq:current}  
{\cal J}^{\mu}\,=\,\sum_{ij}{\overline
\psi}_i\gamma^{\mu}(1-\gamma^5)V_{ij}\psi_j\,, 
\end{equation} 
stands for the usual $V-A$ weak current, $V_{ij}$ denotes the
Cabibbo-Kobayashi-Maskawa flavor-mixing matrix elements, $\psi_i$
represents a light quark flavor, $u$, $d$, or $s$, and finally,
$G_F$ is the so-called Fermi coupling constant. Let $\Delta Q$,
$\Delta S$, and $\Delta I$ denote respectively the change of
charge, strangeness, and isospin of the current (\ref{eq:current}).
Since the latter satisfies the $\Delta S=\Delta Q$ rule, then,
three physical modes for the decay (\ref{eq:processes}) are
possible~\cite{Okun:1959pe, Chadan:1959on}, 
\begin{eqnarray} 
K^+(p)
&\longrightarrow & \pi^+(p_1)\,\pi^-(p_2)\,\ell^+(p_l)\,\nu_{\ell}(p_{\nu})\,, 
\label{eq:charged} \\
K^+(p)
&\longrightarrow & \pi^0(p_1)\,\pi^0(p_2)\,\ell^+(p_l)\,\nu_{\ell}(p_{\nu})\,, 
\label{eq:neutral} \\
K^0(p)
&\longrightarrow & \pi^0(p_1)\,\pi^-(p_2)\,\ell^+(p_l)\,\nu_{\ell}(p_{\nu})\,. 
\label{eq:mixed} 
\end{eqnarray} 
The corresponding decay amplitudes will be denoted by ${\cal A}^{+-}$, ${\cal A}^{00}$, and ${\cal A}^{0-}$, respectively. From now on, we will refer to the decay (\ref{eq:charged}) as the \textit{charged channel}, the decay (\ref{eq:neutral}) as the \textit{neutral channel}, and the decay (\ref{eq:mixed}) as the \textit{mixed channel}.

\subsection{Isospin decomposition}

Pions and kaons are isospin eigenstates with eigenvalues $1$ and
$1/2$, respectively. It follows that the two pions in the final
state of (\ref{eq:processes}) have a total isospin $I=0,1,2$. Assuming
that the current (\ref{eq:current}) satisfies the $\Delta I=1/2$ rule,
then, the two-pion system must be a state of isospin $0$ and $1$
only. Moreover, one can assimilate the current (\ref{eq:current}) to a
spurion carrying an isospin $1/2$ and think about the strong
interaction part of $K_{\ell 4}$ decays as being a scattering between
the spurion-kaon system and the pion-pion one. Accordingly, the
decay amplitudes for the reactions (\ref{eq:charged})-(\ref{eq:mixed})
can be expressed in terms of the invariant amplitudes, ${\cal
A}^0$ and ${\cal A}^1$, for $K_{\ell 4}$ transitions to states of
definite isospin, $I=0$ and $I=1$, respectively~\footnote{We use
Condon-Shortley phase conventions.}, 
\begin{eqnarray} 
{\cal A}^{+-}
&=& \frac{1}{2}\,{\cal A}^1-\frac{1}{\sqrt{6}}\,{\cal A}^0\,,  
\\
{\cal A}^{00}
&=& \frac{1}{\sqrt{6}}\,{\cal A}^0\,,  
\\
{\cal A}^{0-}
&=& \frac{1}{\sqrt{2}}\,{\cal A}^1\,. 
\end{eqnarray} 
This means that the amplitudes for the various processes are
related, 
\begin{equation} 
\label{eq:isospin_relation} 
{\cal A}^{+-}\,=\,\frac{1}{\sqrt{2}}\,{\cal A}^{0-}-{\cal A}^{00}\,. 
\end{equation}
The Bose symmetry of the two-pion system reflects an invariance under the interchange of the two pions in the final state, $p_1\leftrightarrow p_2$. Therefore, the isospin wave function of the two-pion system is symmetric for $I=0$ and antisymmetric for $I=1$ under the interchange, $p_1\leftrightarrow p_2$. Moreover, in accordance with Bose principle, the orbital angular momentum of the two-pion system should be even for $I=0$ and odd for $I=1$.

\subsection{Form factors}

To see the consequences of the foregoing on the dynamics of $K_{\ell 4}$
decays, let us consider the charged process, (\ref{eq:charged}), and
introduce the notations, 
\begin{equation}
P\,=\,p_1+p_2\,, \;
Q\,=\,p_1-p_2\,, \; L\,=\,p_l+p_{\nu}\,, \;
N\,=\,p_l-p_{\nu}\,. 
\end{equation} 
The decay amplitude for the process in question can be written from the foregoing as
follows, 
\begin{eqnarray} 
{\cal A}^{+-}
&=& i\,\frac{G_F}{\sqrt{2}}\,V_{us}^*{\overline u}(\boldsymbol{p}_{\nu})\gamma_{\mu}(1-\gamma^5)v(\boldsymbol{p}_l)\times 
\nonumber \\
&& \qquad \langle\pi^+(p_1)\pi^-(p_2)|{\overline s}\gamma^{\mu}(1-\gamma^5)u|K^+(p)\rangle\,. 
\end{eqnarray} 
Due to the opposit relative intrinsic parities of the $K^+$
and $\pi^+\pi^-$ states, the matrix element of the vector current
between these two states transforms as an axial vector, whereas
that of the axial current transforms as a vector. Therefore, the
hadronic matrix element appearing in the preceding equation
possesses the following Lorentz decomposition, 
\begin{eqnarray} 
&& \langle\pi^+(p_1)\pi^-(p_2)|{\overline
s}\gamma^{\mu}u|K^+(p)\rangle 
\nonumber \\ 
&& \qquad\qquad \doteq\,-\frac{1}{M_{K^{\pm}}^3}\,\epsilon^{\mu\nu\rho\sigma}
L_{\nu}P_{\rho}Q_{\sigma}\,H^{+-}\,,
\label{eq:vector_current} \\
&& \langle\pi^+(p_1)\pi^-(p_2)|{\overline
s}\gamma^{\mu}\gamma^5u|K^+(p)\rangle 
\nonumber \\ 
&& \qquad\qquad \doteq\,-\frac{i}{M_{K^{\pm}}}\,\left
(\,P^{\mu}F^{+-}+Q^{\mu}G^{+-}+L^{\mu}R^{+-}\right )\,, 
\label{eq:axial_vector_current} 
\end{eqnarray} 
The $K_{\ell 4}$ form factors for the charged decay, $H^{+-}$,
$F^{+-}$, $G^{+-}$, and $R^{+-}$ are analytic functions of three
independent Lorentz invariants which we denote by, 
\begin{equation}
s_{\pi}\,=\,P^2\,, \quad t_{\pi}\,=\,(p-p_1)^2\,,
\quad u_{\pi}\,=\,(p-p_2)^2\,. 
\end{equation} 
They are made dimensionless by inserting the normalizations, $M_{K^{\pm}}^{-3}$ and $M_{K^{\pm}}^{-1}$, in (\ref{eq:vector_current}) and (\ref{eq:axial_vector_current}), respectively. The fact that we have used the \textit{charged} kaon mass is a purely conventional matter and corresponds to the choice of defining the isospin limit
in terms of charged masses. 

The interchange $p_1\leftrightarrow p_2$ is equivalent to $t_{\pi}\leftrightarrow u_{\pi}$. It is then convenient to introduce the combinations, 
\begin{eqnarray} 
2I_{\pm}^{+-}
&\doteq & I^{+-}(s_{\pi},t_{\pi},u_{\pi})\pm I^{+-}(s_{\pi},u_{\pi},t_{\pi})\,, 
\nonumber \\
I
&\doteq & H\,,\,F\,,\,G\,,\,R\,, 
\end{eqnarray} 
where the $I^{+-}(s_{\pi},t_{\pi},u_{\pi})$ are defined in (\ref{eq:vector_current}) and (\ref{eq:axial_vector_current}). Similar notations hold for the neutral and mixed channels. From the foregoing, the decay amplitude ${\cal A}^{00}$ is symmetric under $t_{\pi}\leftrightarrow u_{\pi}$, whereas ${\cal A}^{0-}$ is antisymmetric. Thus, the corresponding form factors satisfy, 
\begin{equation}
H_+^{00}\,=\,F_-^{00}\,=\,G_+^{00}\,=\,R_-^{00}\,=\,0\,, \; 
H_-^{0-}\,=\,F_+^{0-}\,=\,G_-^{0-}\,=\,R_+^{0-}\,=\,0\,.
\end{equation} 
Since (\ref{eq:isospin_relation}) is also satisfied by
the form factors, one obtains the following relations between the
non vanishing parts in form factors of processes (\ref{eq:neutral})
and (\ref{eq:mixed}), 
\begin{equation}
\left (H_-\,,\,F_+\,,\,G_-\,,\,R_+\right )^{00} \,=\,-\left (H_-\,,\,F_+\,,\,G_-\,,\,R_+\right )^{+-}\,, 
\end{equation}
\begin{equation}
\left (H_+\,,\,F_-\,,\,G_+\,,\,R_-\right )^{0-}\,=\,\sqrt{2}\left (H_+\,,\,F_-\,,\,G_+\,,\,R_-\right )^{+-}\,.
\end{equation} 
Hence, it is sufficient to study one of the processes (\ref{eq:charged})-(\ref{eq:mixed}) in the isospin limit.

l
\section{Isospin breaking}
\label{sec:isospin_breaking} 

Isospin breaking affects the decay rate as follows. It generates corrections to the form factors $F$, $G$, $R$ and $H$. It introduces a supplementary form factor $T$. It modifies the Dalitz plot and contributes to the intensity spectrum. 

\subsection{Kinematics}
\label{sec:Kinematics}

\begin{figure}[p]
\epsfxsize14cm \centerline{\epsffile{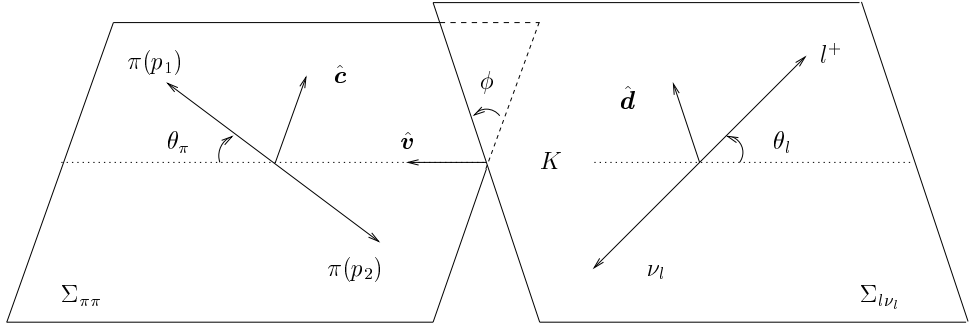}}
\caption{\label{fig:kinematics} Angles and unit vectors used in the
kinematical description of $K_{\ell 4}$ decays. $\Sigma_{\pi\pi}$ and
$\Sigma_{l\nu_l}$ are the planes defined in the kaon rest frame by
the pion pair and the lepton pair, respectively. $\theta_{\pi}$
($\theta_l$), the angle formed by $\boldsymbol{p}_1$
($\boldsymbol{p}_l$), in the dipion (dilepton) rest frame, and the
line of flight of the dipion (dilepton) as defined in the kaon
rest frame. $\phi$, the angle between the normals to
$\Sigma_{\pi\pi}$ and $\Sigma_{l\nu_l}$. $\hat{\boldsymbol{v}}$ is
a unit vector along the direction of flight of the dipion in the
kaon rest frame. $\hat{\boldsymbol{c}}$ ($\hat{\boldsymbol{d}}$)
is a unit vector along the projection of $\boldsymbol{p}_1$
($\boldsymbol{p}_l$) perpendicular to $\hat{\boldsymbol{v}}$.}
\end{figure}

We follow the approach used in~\cite{Cabibbo:1965} to study the
kinematics of $K_{\ell 4}$ decays. The idea consists on looking at these
decays as being two-body decays into a dipion of mass $s_{\pi}$
and a dilepton of mass $s_l\,\doteq\,L^2$. The two systems subsequently
decay in their own center-of-mass frames. To describe the decay
distribution it is convenient to use, besides the invariant
masses, $s_{\pi}$ and $s_l$, the angles $\theta_{\pi}$,
$\theta_l$, and $\phi$ as illustrated in Fig.~\ref{fig:kinematics}.

All of the scalar products obtained from momenta, $p_1$, $p_2$,
$p_l$, and $p_{\nu}$, or equivalently from, $P$, $Q$, $L$, and
$N$, can be expressed in terms of the five independent variables,
$s_{\pi}$, $s_l$, $\theta_{\pi}$, $\theta_l$, and $\phi$. In the kaon rest frame, the Dirac components of momenta read,
\begin{eqnarray}
p^0
&=& M\,, 
\\ 
p^1
&=& p^2\,=\,p^3\,=\,0\,, 
\\ 
p_1^0
&=& \dfrac{1}{4Ms_{\pi}}\left[ (M^2+s_{\pi}-s_l)(s_{\pi}+M_1^2-M_2^2)\right. 
\nonumber \\ 
&& \left.+\lambda^{1/2}(M^2,s_{\pi},s_l)\lambda^{1/2}(s_{\pi},M_1^2,M_2^2)
\cos\theta_{\pi}\right] \,, 
\\ 
p_1^1
&=& \dfrac{1}{4Ms_{\pi}}\left[ (s_{\pi}+M_1^2-M_2^2)\lambda^{1/2}(M^2,s_{\pi},s_l)\right. 
\nonumber \\ 
&& \left. +(M^2+s_{\pi}-s_l)\lambda^{1/2}(s_{\pi},M_1^2,M_2^2)
\cos\theta_{\pi}\right] \,, 
\\ 
p_1^2
&=& \dfrac{1}{2\sqrt{s_{\pi}}}\,\lambda^{1/2}(s_{\pi},M_1^2,M_2^2)\sin\theta_{\pi}\,, 
\\ 
p_1^3
&=& 0\,, 
\\ 
p_2^0
&=& \dfrac{1}{4Ms_{\pi}}\left[ (M^2+s_{\pi}-s_l)(s_{\pi}-M_1^2+M_2^2)\right. 
\nonumber \\ 
&& \left. -\lambda^{1/2}(M^2,s_{\pi},s_l)\lambda^{1/2}(s_{\pi},M_1^2,M_2^2)
\cos\theta_{\pi}\right] \,, 
\\ 
p_2^1
&=& \dfrac{1}{4Ms_{\pi}}\left[ (s_{\pi}-M_1^2+M_2^2)\lambda^{1/2}(M^2,s_{\pi},s_l)\right. 
\nonumber \\ 
&& \left. -(M^2+s_{\pi}-s_l)\lambda^{1/2}(s_{\pi},M_1^2,M_2^2)
\cos\theta_{\pi}\right] \,, 
\\ 
p_2^2
&=& -\dfrac{1}{2\sqrt{s_{\pi}}}\,\lambda^{1/2}(s_{\pi},M_1^2,M_2^2)\sin\theta_{\pi}\,, 
\\ 
p_2^3
&=& 0\,, 
\\ 
p_l^0
&=& \dfrac{1}{4Ms_l}\left[ (s_l+m_l^2)(M^2+s_l-s_{\pi})\right. 
\nonumber \\ 
&& \left. +(s_l-m_l^2)\lambda^{1/2}(M^2,s_{\pi},s_l)\cos\theta_l\right] \,, 
\\ 
p_l^1
&=& -\dfrac{1}{4Ms_l}\left[ (s_l+m_l^2)\lambda^{1/2}(M^2,s_{\pi},s_l)\right. 
\nonumber \\ 
&& \left. +(s_l-m_l^2)(M^2-s_{\pi}+s_l)\cos\theta_l\right] \,, 
\\
p_l^2
&=& \dfrac{1}{2\sqrt{s_l}}\,(s_l-m_l^2)\sin\theta_l\cos\phi \,, 
\\
p_l^3
&=& -\dfrac{1}{2\sqrt{s_l}}\,(s_l-m_l^2)\sin\theta_l\sin\phi \,, 
\\ 
p_{\nu}^0
&=& \dfrac{1}{4Ms_l}\,(s_l-m_l^2)\left[ M^2-s_{\pi}+s_l-\lambda^{1/2}(M^2,s_{\pi},s_l)
\cos\theta_l\right] \,, 
\\ 
p_{\nu}^1
&=& -\dfrac{1}{4Ms_l}\,(s_l-m_l^2)\left[ \lambda^{1/2}(M^2,s_{\pi},s_l)-(M^2-s_{\pi}+s_l)
\cos\theta_l\right] \,, 
\\
p_{\nu}^2
&=& -\dfrac{1}{2\sqrt{s_l}}\,(s_l-m_l^2)\sin\theta_l\cos\phi \,, 
\\
p_{\nu}^3
&=& \dfrac{1}{2\sqrt{s_l}}\,(s_l-m_l^2)\sin\theta_l\sin\phi \,.  
\end{eqnarray} 
In the preceding equations, our choice for the metric tensor is, 
$$
[\eta ]\,=\,\mathrm{diag}\{1,-1,-1,-1\}\,.
$$ 
Therefore, it is easy to obtain the scalar products,
\begin{eqnarray}
P^2
&=& s_{\pi}\,, 
\\
Q^2
&=& \frac{\Delta_{12}^2}{s_{\pi}}-s_{\pi}Y^2\,, 
\\
L^2
&=& s_l\,, 
\\
N^2
&=& 2m_l^2-s_l\,, 
\\
P\cdot Q
&=& \Delta_{12}\,, 
\\
P\cdot L
&=& \frac{1}{2}\,(M^2-s_{\pi}-s_l)\,, 
\\
P\cdot N
&=& \frac{1}{2}\,(M^2-s_{\pi}-s_l)\,z_l+(1-z_l)X\cos\theta_l\,,
\\
Q\cdot L
&=& (M^2-s_{\pi}-s_l)\,\frac{\Delta_{12}}{2s_{\pi}}+XY\cos\theta_{\pi}\,,
\\
Q\cdot N
&=& (1-z_l)X\,\frac{\Delta_{12}}{s_{\pi}}\,\cos\theta_l
        +(M^2-s_{\pi}-s_l)\,\frac{z_l}{2}\,\frac{\Delta_{12}}{s_{\pi}} 
\nonumber \\ 
&& +z_lXY\cos\theta_{\pi}+\frac{1}{2}\,(1-z_l)Y\,[\,(M^2-s_{\pi}-s_l)\times
\nonumber \\ 
&& \cos\theta_{\pi}\cos\theta_l
-2\sqrt{s_{\pi}s_l}\sin\theta_{\pi}\sin\theta_l\cos\phi\,]\,,
\\
L\cdot N
&=& m_l^2\,, 
\\
\epsilon_{\mu\nu\rho\sigma}L^{\mu}N^{\nu}P^{\rho}Q^{\sigma}
&=& -(1-z_l)XY\sqrt{s_{\pi}s_l}\sin\theta_{\pi}\sin\theta_l\sin\phi\,,
\end{eqnarray} 
with the following notations,
\begin{eqnarray}
\Delta_{12}
&\doteq& M_1^2-M_2^2\,, 
\\
z_l
&\doteq& \frac{m_l^2}{s_l}\,, 
\\
X
&\doteq& \frac{1}{2}\,\lambda^{1/2}(M^2,s_{\pi},s_l)\,, 
\\
Y
&\doteq& \frac{1}{s_{\pi}}\,\lambda^{1/2}(s_{\pi},M_1^2,M_2^2)\,.
\end{eqnarray}
The function,
\begin{equation}
\lambda (x,y,z)\,\doteq\,x^2+y^2+z^2-2xy-2xz-2yz\,,
\end{equation}
is the usual K\"all\'en function.

\subsection{Phase space integral}
\label{sec:phase_space_integral}

The starting point is the $K_{\ell 4}$ differential phase space,
\begin{eqnarray}
d\Phi
&=& (2\pi )^4\delta^{(4)}(p_1+p_2+p_l+p_{\nu}-p)\times 
\nonumber \\
&& \frac{d^3\boldsymbol{p}_1}{(2\pi )^32E_1}\,
        \frac{d^3\boldsymbol{p}_2}{(2\pi )^32E_2}\,
        \frac{d^3\boldsymbol{p}_l}{(2\pi )^32E_l}\,
        \frac{d^3\boldsymbol{p}_{\nu}}{(2\pi
        )^32|\boldsymbol{p}_{\nu}|}\,,
\nonumber
\end{eqnarray}
with particle energies,
\begin{equation}
E_i\,=\,\sqrt{p_i^2+\boldsymbol{p}_i^2}\,.
\end{equation}
This can be written as
\begin{eqnarray}
d\Phi
&=& \int d^4P\int d^4L\,(2\pi )^4\delta^{(4)}(P+L-p)\times 
\nonumber \\
&& \frac{d^3\boldsymbol{p}_1}{(2\pi )^32E_1}\,
        \frac{d^3\boldsymbol{p}_2}{(2\pi )^32E_2}\,
        \delta^{(4)}(p_1+p_2-P)\times 
\nonumber \\
&& \frac{d^3\boldsymbol{p}_l}{(2\pi )^32E_l}\,
        \frac{d^3\boldsymbol{p}_{\nu}}{(2\pi )^32|\boldsymbol{p}_{\nu}|}
        \delta^{(4)}(p_l+p_{\nu}-L)\,.
\nonumber
\end{eqnarray}
Performing the integrations over the $\delta$ functions
separately, we obtain for the phase space,
$$
\int d\Phi\,=\,\frac{1}{4M^2}\,(2\pi )^5\int ds_{\pi}\int
ds_l\,\lambda^{1/2}(M^2,s_{\pi},s_l)\,\Phi_{\pi\pi}\,\Phi_{l\nu_l}\,,
$$
with the dipion phase space,
\begin{eqnarray}
\Phi_{\pi\pi}
&=& \int\frac{d^3\boldsymbol{p}_1}{(2\pi )^32E_1}\,
        \frac{d^3\boldsymbol{p}_2}{(2\pi )^32E_2}\,
        \delta^{(4)}(p_1+p_2-P) 
\nonumber \\
&=& \frac{1}{(2\pi )^5}\,\frac{1}{8s_{\pi}}\,
        \lambda^{1/2}(s_{\pi},M_1^2,M_2^2)\int
        d(\cos\theta_{\pi})\,,
\end{eqnarray} 
and the dilepton phase space,
\begin{eqnarray}
\Phi_{l\nu_l}
&=& \int\frac{d^3\boldsymbol{p}_l}{(2\pi )^32E_l}\,
        \frac{d^3\boldsymbol{p}_{\nu}}{(2\pi )^32|\boldsymbol{p}_{\nu}|}
        \delta^{(4)}(p_l+p_{\nu}-L) 
\nonumber \\
&=& \frac{1}{(2\pi )^6}\,\frac{1}{8}\,(1-z_l)
        \int d\phi\int d(\cos\theta_l)\,.
\end{eqnarray}
From the foregoing, the $K_{\ell 4}$ phase space can be obtained by
integrating the differential phase space,
$$
d^5\Phi\,=\,M^3N(s_{\pi},s_l)ds_{\pi}ds_l
d(\cos\theta_{\pi})d(\cos\theta_l)d\phi\,,
$$
with,
$$
N(s_{\pi},s_l)\,\doteq\,\frac{1}{2^{13}\pi^6}\,\frac{1}{M^5}\,(1-z_l)XY\,,
$$
over the following range of variables,
\begin{equation}
\begin{array}{ccccc}
0              & \leq & \phi         & \leq & 2\pi \\
0              & \leq & \theta_l     & \leq & \pi \\
0              & \leq & \theta_{\pi} & \leq & \pi \\
m_l^2          & \leq & s_l          & \leq & (M-\sqrt{s_{\pi}})^2 \\
2(M_1^2+M_2^2) & \leq & s_{\pi}      & \leq & (M-m_l)^2\,.
\end{array}
\end{equation}

\subsection{The decay rate}
\label{sec:decay_rate}

Electromagnetism breaks the $V-A$ structure of $K_{\ell 4}$ decay
amplitudes. By Lorentz covariance, these can be written in general
as follows,
\begin{eqnarray}
{\cal A}
&\doteq & \frac{G_FV_{us}^*}{\sqrt{2}}\,{\overline u}(\boldsymbol{p}_{\nu})
        (1+\gamma^5)\left
        [\frac{1}{M_{K^{\pm}}}\,(fP^{\mu}+gQ^{\mu}+rL^{\mu})\gamma^{\mu}\right.
\nonumber \\
&+& \left. i\,\frac{h}{M_{K^{\pm}}^3}\,\epsilon_{\mu\nu\rho\sigma}
        \gamma^{\mu}L^{\nu}P^{\rho}Q^{\sigma}
        -i\,\frac{T}{M_{K^{\pm}}^2}\,\sigma_{\mu\nu}\,p_1^{\mu}p_2^{\nu}\right
        ]v(\boldsymbol{p}_l)\,, 
\nonumber
\end{eqnarray}
where,
$$
\sigma^{\mu\nu}\,\doteq\,\dfrac{i}{2}\left[ \gamma^{\mu},\gamma^{\nu}\right] \,.
$$ 
The quantities $f$, $g$, $r$, and $h$, will be called the
\textit{corrected} $K_{\ell 4}$ form factors since their isospin
limits are nothing else than the $K_{\ell 4}$ form factors, $F$, $G$,
$R$, and $H$, respectively. The normalization factors are written
in powers of the charged kaon mass by convention (this corresponds
to our definition of the isospin limit as stated in
Sec.~\ref{sec:isospin_limit}). Squaring the amplitude and summing
over spins, we get,
$$
\sum_{\mathrm{spins}}|{\cal A}|^2\,=\,2G_F^2|V_{us}|^2\,\frac{1}{M^2}\,j_5(s_{\pi},s_l,\theta_{\pi},\theta_l,\phi
)\,,
$$
with the following expression for the intensity spectrum,
\begin{eqnarray}
j_5
&=& |f|^2\left[ (P\cdot L)^2-(P\cdot N)^2-s_{\pi}(s_l-m_l^2)\right] 
\nonumber \\ 
&+& |g|^2\left[ (Q\cdot L)^2-(Q\cdot N)^2-Q^2(s_l-m_l^2)\right] 
\nonumber \\ 
&+& |r|^2m_l^2(s_l-m_l^2) 
\nonumber \\ 
&-& |h|^2\,\dfrac{1}{M^4}\left\lbrace (\epsilon_{\mu\nu\rho\sigma}L^{\mu}N^{\nu}P^{\rho}Q^{\sigma})^2+(s_l-m_l^2)\left[ Q^2X^2+s_{\pi}(Q\cdot L)^2\right.\right. 
\nonumber \\ 
&+& \left.\left. s_l(P\cdot Q)^2-2(P\cdot Q)(Q\cdot L)(P\cdot L)\right] \right\rbrace  
\nonumber \\ 
&+& (f^*g+fg^*)\left[ (P\cdot L)(Q\cdot L)-(P\cdot N)(Q\cdot N)-(s_l-m_l^2)(P\cdot Q)\right] 
\nonumber \\ 
&+& (f^*r+fr^*)m_l^2\left[ (P\cdot L)-(P\cdot N)\right] 
\nonumber \\ 
&+& (f^*h+fh^*)\,\dfrac{1}{M^2}\left[ (Q\cdot N)(P\cdot L)^2-(Q\cdot L)(P\cdot L)(P\cdot N)\right. 
\nonumber \\ 
&-& \left. s_{\pi}s_l(Q\cdot N)+m_l^2s_{\pi}(Q\cdot L)-m_l^2(P\cdot Q)(P\cdot L)+s_l(P\cdot Q)(P\cdot N)\right] 
\nonumber \\ 
&+& (g^*r+gr^*)m_l^2\left[ (Q\cdot L)-(Q\cdot N)\right] 
\nonumber \\ 
&+& (g^*h+gh^*)\,\dfrac{1}{M^2}\left[ (P\cdot L)(Q\cdot L)(Q\cdot N)-(P\cdot N)(Q\cdot L)^2\right. 
\nonumber \\ 
&+& \left. s_lQ^2(P\cdot N)-m_l^2Q^2(P\cdot L)+m_l^2(P\cdot Q)(Q\cdot L)-s_l(P\cdot Q)(Q\cdot N)\right] 
\nonumber \\ 
&-& \epsilon_{\mu\nu\rho\sigma}L^{\mu}N^{\nu}P^{\rho}Q^{\sigma}\,\dfrac{i}{M^2}\left[ M^2(f^*g-fg^*)\right. 
\nonumber \\ 
&-& \left.(P\cdot N)(f^*h-fh^*)-(Q\cdot N)(g^*h-gh^*)-m_l^2(r^*h-rh^*)\right] 
\nonumber \\
&+& (fT^*+Tf^*)\,\dfrac{m_l}{2M}\left[ s_{\pi}(Q\cdot N)\right. 
\nonumber \\ 
&-& \left. s_{\pi}(Q\cdot L)+(P\cdot Q)(P\cdot L)-(P\cdot Q)(P\cdot N)\right] 
\nonumber \\ 
&+& (gT^*+Tg^*)\,\dfrac{m_l}{2M}\left[ Q^2(P\cdot L)\right. 
\nonumber \\
&-& \left. Q^2(P\cdot N)-(P\cdot Q)(Q\cdot L)+(P\cdot Q)(Q\cdot N)\right] 
\nonumber \\ 
&+& (rT^*+Tr^*)\,\dfrac{m_l}{2M}\left[ (P\cdot L)(Q\cdot N)-(P\cdot N)(Q\cdot L)\right] 
\nonumber \\ 
&+& (rT^*-Tr^*)\,\dfrac{im_l}{2M}\,\epsilon_{\mu\nu\rho\sigma}
L^{\mu}N^{\nu}P^{\rho}Q^{\sigma} 
\nonumber \\ 
&-& (hT^*+Th^*)\,\dfrac{m_l}{2M^3}\left[ 2(P\cdot Q)(P\cdot L)(Q\cdot L)-(P\cdot Q)(P\cdot L)(Q\cdot N)\right. 
\nonumber \\ 
&-& Q^2(P\cdot L)^2+Q^2(P\cdot L)(P\cdot N)-s_{\pi}(Q\cdot L)^2
\nonumber \\ 
&+& s_{\pi}(Q\cdot L)(Q\cdot N)-(P\cdot Q)(P\cdot N)(Q\cdot L)-s_l(P\cdot Q)^2
\nonumber \\ 
&+& \left. (P\cdot Q)^2(L\cdot N)+s_{\pi}s_lQ^2-s_{\pi}Q^2(L\cdot N)\right] 
\nonumber \\ 
&+& |T|^2\,\dfrac{1}{8M^2}\left[ s_{\pi}s_lQ^2-s_l(P\cdot Q)^2-2Q^2(P\cdot L)^2\right. 
\nonumber \\ 
&+& 4(P\cdot Q)(P\cdot L)(Q\cdot L)-2s_{\pi}(Q\cdot L)^2-s_{\pi}Q^2N^2+(P\cdot Q)^2N^2
\nonumber \\ 
&+& \left. 2Q^2(P\cdot N)^2+2s_{\pi}(Q\cdot N)^2-4(P\cdot Q)(P\cdot N)(Q\cdot N)\right] \,.
\end{eqnarray} 
The $K_{\ell 4}$ differential decay rate is defined by,
$$
d\Gamma\,\doteq\,\frac{1}{2M}\,d\Phi\sum_{\mathrm{spins}}|{\cal A}|^2\,.
$$
It reads, in terms of the five independent variables,
\begin{equation}
d^5\Gamma\,=\,G_F^2|V_{us}|^2N(s_{\pi},s_l)j_5(s_{\pi},s_l,\theta_{\pi},\theta_l,\phi
)ds_{\pi}ds_ld(\cos\theta_{\pi})d(\cos\theta_l)d\phi\,.
\end{equation}

\subsection{The effective Lagrangian}

In order to treat completely electromagnetic effects in $K_{\ell 4}$
decays, not only the pseudoscalars but also the photon
and the light leptons have to be included as dynamical degrees of freedom
in an appropriate effective Lagrangian~\cite{Knecht:1999ag}. The starting point is QCD in the limit $m_u = m_d = m_s = 0$. The resulting chiral symmetry, $G = SU(3)_L \times SU(3)_R$, is spontaneously broken to $SU(3)_V$. The pseudoscalar mesons $(\pi ,K,\eta )$ are nothing else than the corresponding Goldstone fields $\phi_i$ ($i = 1,\ldots ,8$) acting as coordinates of the coset space $SU(3)_L \times SU(3)_R/SU(3)_V$.
The transformation rules for the coset variables $u_{L,R}(\phi )$ are
\begin{eqnarray}
u_L(\phi ) 
&\stackrel{G}{\rightarrow}& g_L u_L h(g,\phi )^{-1}\,, 
\nonumber \\
u_R(\phi) 
&\stackrel{G}{\rightarrow}& g_R u_R h(g,\phi)^{-1}\,, 
\nonumber \\
g = (g_L,g_R) 
&\in & SU(3)_L \times SU(3)_R\,,
\end{eqnarray}
where $h(g,\phi )$ is the nonlinear realization of $G$~\cite{Coleman:1969sm, Callan:1969sn}.

As stated before, the photon field $A_{\mu}$ and the leptons $\ell ,\nu_{\ell}$ ($\ell =\mathrm{e},\mu$) have to be dynamical. Thus, they most be introduced in 
the covariant derivative,
\begin{equation}
u_{\mu}\,\doteq\,i\,[\,u_R^{\dagger}(\partial_{\mu}-ir_{\mu})u_R-u_L^{\dagger}
(\partial_{\mu}-il_{\mu})u_L\,]\,,
\end{equation}
by adding appropriate terms to the usual external vector and axial-vector
sources ${\cal V}_{\mu}$, ${\cal A}_{\mu}$. At the quark level, this procedure corresponds to the usual minimal coupling prescription in the case of electromagnetism, and to Cabibbo universality in the case of the charged weak currents,
\begin{eqnarray} 
\label{sources}
l_{\mu} 
&\doteq & v_{\mu}-a_{\mu}-eQ_L^{\mathrm{em}}A_{\mu}+\sum_{\ell}
(\overline{\ell}\gamma_{\mu}\nu_{\ell L}Q_L^{\mathrm{w}}+\overline{\nu_{\ell L}}
\gamma_{\mu} \ell Q_L^{{\mathrm{w}}\dagger})\,, 
\nonumber \\
r_{\mu} 
&\doteq & v_{\mu}+a_{\mu}-eQ_R^{\mathrm{em}}A_{\mu}\,.
\end{eqnarray}
The $3 \times 3$ matrices $Q_{L,R}^{\mathrm{em}}$, $Q_L^{\mathrm{w}}$ are
spurion fields corresponding to electromagnetic and weak coupling, respectively. They transform as,
\begin{equation}
Q_L^{\mathrm{em,w}} \stackrel{G}{\rightarrow} g_L Q_L^{\mathrm{em,w}} g_L^{\dagger}\,, \qquad
Q_R^{\mathrm{em}} \stackrel{G}{\rightarrow} g_R Q_R^{\mathrm{em}} g_R^{\dagger}\,,
\end{equation}
under the chiral group. In practical calculations, one identifies $Q_{L,R}^{\mathrm{em}}$ with the quark charge matrix
\begin{equation} 
\label{Qem}
Q^{\mathrm{em}}\,\doteq\,\left ( \begin{array}{ccc} 
                            2/3 & 0    & 0 \\ 
                              0 & -1/3 & 0 \\ 
                              0 & 0    & -1/3 
                            \end{array}\right )\,,
\end{equation}
whereas the weak spurion is replaced by,
\begin{equation} 
\label{Qw}
Q_L^{\mathrm{w}}\,\doteq\, -2\sqrt{2}\; G_F \left ( \begin{array}{ccc}
                                               0 & V_{ud} & V_{us} \\ 
                                               0 & 0      & 0 \\ 
                                               0 & 0      & 0 
                                               \end{array} \right )\,,
\end{equation}
where $G_F$ is the Fermi coupling constant and $V_{ud}$, $V_{us}$ are
Kobayashi-Maskawa matrix elements. 

In order to take into account electromagnetic mass difference between charged and neutral mesons, it is convenient to define the following electromagnetic and weak sources,
\begin{equation} 
\label{Qhom}
{\cal Q}_L^{\mathrm{em,w}}\,\doteq\,u_L^{\dagger} Q_L^{\mathrm{em,w}} u_L, \qquad
{\cal Q}_R^{\mathrm{em}}\,\doteq\,u_R^{\dagger} Q_R^{\mathrm{em}} u_R\,
\end{equation}
transforming as
\begin{eqnarray}
{\cal Q}_L^{\mathrm{em,w}} 
&\stackrel{G}{\rightarrow}& h(g,\phi ) {\cal Q}_L^{\mathrm{em,w}} h(g,\phi )^{-1}\,,
\nonumber \\
{\cal Q}_R^{\mathrm{em}} 
&\stackrel{G}{\rightarrow}& h(g,\phi ) {\cal Q}_R^{\mathrm{em}} h(g,\phi )^{-1}\,.
\end{eqnarray}

With these building blocks, the lowest order effective Lagrangian takes
the form
\begin{eqnarray} 
\label{Leff}
{\cal L}_{\mathrm{eff}} 
&=& \frac{{F_0}^2}{4} \; \langle u_{\mu} u^{\mu} + \chi_+\rangle +
e^2 {F_0}^4 Z_0 \langle {\cal Q}_L^{\mathrm{em}} {\cal Q}_R^{\mathrm{em}}\rangle \nonumber \\
&-& \frac{1}{4} F_{\mu\nu} F^{\mu\nu} + \sum_{\ell}
[\overline{\ell}(i \! \not\!\partial + e \! \not\!\!A - m_l)\ell +
\overline{\nu}_{\ell L} \, i \! \not\!\partial \nu_{\ell L}],
\end{eqnarray}
where $\langle \;\rangle$ denotes the trace in three-dimensional flavour
space. 

The low-energy constant $F_0$ appearing in the preceding formula is an order parameter for chiral symmetry since it testifies to its spontaneous breaking. It represents the pion decay constant in the chiral limit, $m_u=m_d=m_s=0$, and in the absence of electroweak interactions. Explicit chiral symmetry breaking due to quark masses is included in,
$$
\chi_+\,\doteq\,u_R^{\dagger} \chi u_L + u_L^{\dagger} \chi^{\dagger} u_R\,.
$$ 
In practice, one makes the following substitution,
\begin{equation}
\chi\,\rightarrow\,2{B_0} {\cal M}_{\mathrm{quark}}\,\doteq\,2B_0\left( 
\begin{array}{ccc}
m_u & 0 & 0 \\
0   & m_d & 0 \\ 
0   &     & m_s 
\end{array}\right)\,, 
\end{equation}
where $B_0$ is an order parameter for chiral symmetry. It is related to the quark condensate in the chiral limit by,
\begin{equation}
\langle\overline{q}q\rangle\,=\,-F_0^2B_0\,.
\end{equation} 
The low-energy constant $Z_0$ expresses explicit chiral symmetry breaking by electromagnetism. It is given by the electromagnetic mass of the pion as we will see below. 

In the absence of electroweak interactions, that is, for $m_u=m_d\,, \; \alpha=0$, ChPT is Isospin-invariant. In order to study Isospin breaking effects in ChPT processes, the usual chiral expansion in powers of $p$ and $m_q$ is no more sufficient. One must also expand matrix elements in powers of the isospin breaking parameters, $m_d-m_u$ and $\alpha$. On the other hand, the best accuracy ever reached in strong interaction observable measurements does not exceed the $5\%$ level. Thus, we consider that an expansion to orders ${\cal O}(m_d-m_u)$ and ${\cal O}(\alpha )$ is highly adequate for our purposes. Moreover, chiral expansion and Isospin breaking expansion have to be related in a consistent way in order to obtain reliable results. We adopt an expansion scheme where the Isospin breaking parameters are considered as quantities of order $p^2$ in the chiral counting, 
\begin{equation}
\label{eq:leading_order} 
{\cal O}(m_d-m_u)\,=\,{\cal O}(\alpha )\,=\,{\cal O}(m_q)\,=\,{\cal O}(p^2)\,.
\end{equation} 
Therefore, tree level calculation corresponding to leading chiral order is caracterized by chiral orders cited in (\ref{eq:leading_order}). Concerning one-loop level calculation which corresponds to next-to-leading chiral order, it is caracterized by the following chiral orders,
$$
\mathcal{O}(p^4)\,,\;\mathcal{O}(m_q^2)\,,\;\mathcal{O}(p^2m_q)\,,
$$
\begin{equation}
\label{eq:next_to_leading_order}
\mathcal{O}\left( p^2(m_d-m_u)\right)\,,\;\mathcal{O}\left( m_q(m_d-m_u)\right)\,,\;\mathcal{O}(p^2\alpha )\,,\;\mathcal{O}(m_q\alpha )\,.
\end{equation}  

We have now all necessary elements to calculate any Green function in the framework of ChPT including Isospin breaking effects. For example, the leading chiral order expressions for light meson masses are found to be,
\begin{eqnarray}
M_{\pi^0}^2
&=& M_{\pi}^2\,, 
\label{eq:neutral_pion} \\ 
M_{\pi^{\pm}}^2
&=& M_{\pi}^2+2Z_0e^2F_0^2\,,
\label{eq:charged_pion} \\
M_{K^0}^2
&=& M_K^2+\dfrac{2\epsilon}{\sqrt{3}}\left( M_K^2-M_{\pi}^2\right) \,,
\label{eq:neutral_kaon} \\
M_{K^{\pm}}^2
&=& M_K^2-\dfrac{2\epsilon}{\sqrt{3}}\left( M_K^2-M_{\pi}^2\right) +2Z_0e^2F_0^2\,,
\label{eq:charged_kaon} \\
M_{\eta}^2
&=& \dfrac{1}{3}\left( 4M_K^2-M_{\pi}^2\right) \,. 
\label{eq:eta_mass} 
\end{eqnarray} 
Herein, $M_{\pi}$ and $M_K$ represent respectively pion and kaon masses in the absence of isospin breaking,
\begin{equation}
\label{eq:quark_masses} 
M_{\pi}^2\,\doteq\,2B_0\hat{m}\,, \quad M_K^2\,\doteq\,B_0(\hat{m}+m_s)\,, \quad 2\hat{m}\,\doteq\,m_u+m_d\,,
\end{equation} 
$\epsilon$ measures the rate of $SU(2)$ to $SU(3)$ breaking,
\begin{equation}
\label{eq:epsilon}
\epsilon\,\doteq\,\dfrac{\sqrt{3}}{4}\,\dfrac{m_d-m_u}{m_s-\hat{m}}\,.
\end{equation}
At next-to-leading chiral order, one-loop calculation is involved. Vertices are extracted from Lagrangian (\ref{Leff}). Meson masses in the propagators as well as in the vertices can be identified with expressions (\ref{eq:neutral_pion})-(\ref{eq:eta_mass}). As is well known, loops are ultraviolet divergent. To remove divergences, renormalization should be employed. The procedure consists on adding to Lagrangian (\ref{Leff}) suitable counter-terms~\cite{Gasser:1985gg, Urech:1995hd, Neufeld:1995eg, Neufeld:1996mu, Knecht:1999ag} generating exactly the same divergences but with opposite sign. Moreover, the cancellation should occur \textit{order by order} in the chiral expansion as dictated by renormalizability principles of effective field theories. Counter-terms are modulated by low-energy constants which are order parameters for chiral symmetry. In order to determine these constants one proceeds as follows. Let $C$ be either a low-energy constant or a combination of low-energy constants and $\Upsilon$ an observable very sensitive to variations of $C$. One first calculate the expression of $\Upsilon$ in the framework of ChPT to any given order and then \textit{match} the obtained expression with an experimental measurement of $\Upsilon$. It is clear that the value for $C$ deduced from this matching does not constitute a \textit{genuine determination} of the low-energy constant. In fact, it represents the value of $C$ at the given chiral order and with the accuracy of the experimental measurement. Note that the method of effective Lagrangian has the disadvantage of an infinitely increasing number of low-energy constants when going to higher and higher orders in the low-energy expansion. For instance, two constants in the strong sector, $B_0$ and $F_0$, and one constant in the electroweak meson sector, $Z_0$, parametrize the leading chiral order. At next-to-leading chiral order, one has ten low-energy constants in the strong sector, $L_1\,, \ldots\,, L_{10}$, fourteen constants in the electroweak meson sector, $K_1\,, \ldots\,, K_{14}$, and seven constants in the electroweak leptonic sector, $X_1\,, \ldots\,, X_7$. The constants, $L_i$, $K_i$ and $X_i$, are divergent. They absorb the divergence of loops via the renormalization, 
\begin{eqnarray}
L_i
&\doteq & L_i^r(\mu )+\Gamma_i\,\overline{\lambda}\,, \qquad i\,=\,1\,,\ldots\,, 10\,, 
\\
K_i
&\doteq & K_i^r(\mu )+\Sigma_i\,\overline{\lambda}\,, \qquad i\,=\,1\,,\ldots\,, 14\,, 
\\
X_i
&\doteq & X_i^r(\mu )+\Xi_i\,\overline{\lambda}\,, \qquad i\,=\,1\,,\ldots\,, 7\,.
\end{eqnarray}
Herein, $\overline{\lambda}$ corresponds to pole subtraction in the $\overline{\mathrm{MS}}$ dimensional regularization scheme (see appendix). The beta-functions, $\Gamma_i$, $\Sigma_i$ and $\Xi_i$, can be found in~\cite{Gasser:1985gg}, \cite{Urech:1995hd} and~\cite{Knecht:1999ag}, respectively. The scale $\mu$ cancels in observables as can be seen from the renormalization group equations,
\begin{equation}
L_i^r(\mu_2)\,=\,L_i^r(\mu_1)+\dfrac{\Gamma_i}{16\pi^2}\,\ln\dfrac{\mu_1}{\mu_2}\,,
\end{equation}
and similar for $K_i$ and $X_i$.   

\section{Leading order}
\label{sec:leading_order}

The charged, neutral, and mixed channels will be studied at tree
level including isospin breaking effects. Due to the absence of a
Lagrangian treating isospin breaking in the anomaly sector~\footnote{See~\cite{Ananthanarayan:2002kj} for the two-flavor case.}, the
corrected form factor $h$ will be put to its isospin limit, $H$.
In the following, we shall give a detailed calculation for the
charged channel due to the presence of a tensorial form factor already at leading order. For the remaining channels, we limit ourselves to quoting results.

\subsection{Form factors}

\begin{figure*}[p]
\epsfxsize14cm \centerline{\epsffile{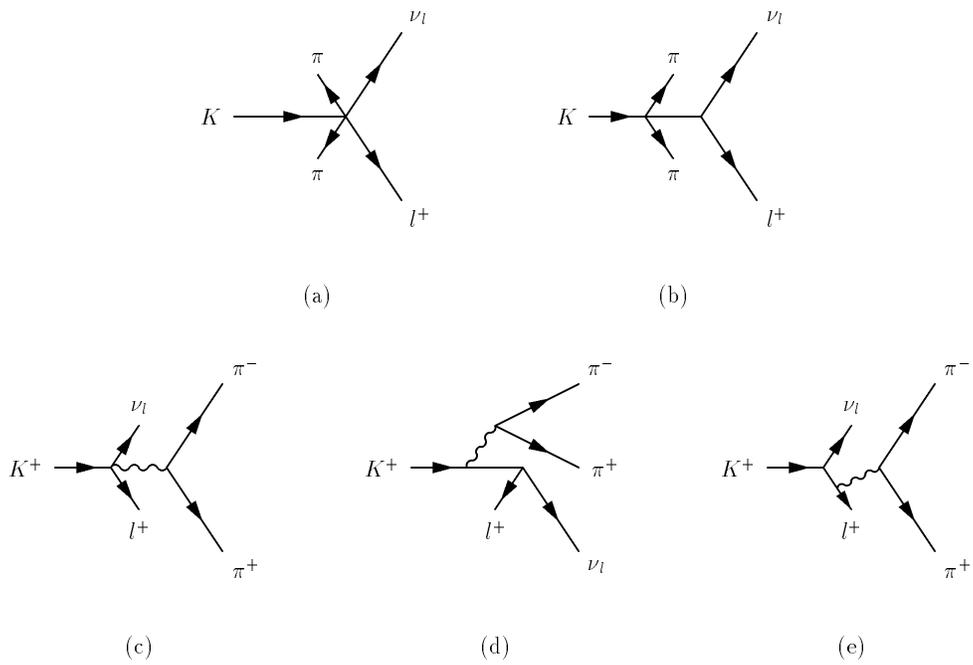}}
\caption{\label{fig:1} Feynman diagrams representing $K_{\ell 4}$ decay
amplitudes at tree level. Wavy lines stand for photons. Only
diagrams (a) and (b) contribute to the decay amplitudes ${\cal
A}^{00}$ and ${\cal A}^{0-}$.}
\end{figure*}

The different topologies for $K_{\ell 4}$ decays at tree level in
perturbation theory are drawn in Fig.~\ref{fig:1}.

The contribution of each Feynman diagram to the charged decay
amplitude will be given separately.

\textbf{diagram 2. (a)}
$$
{\cal A}^{+-}\,=\,\frac{1}{6F_0}\,G_FV_{us}^*{\overline
u}(\boldsymbol{p}_{\nu})(1+\gamma^5)\gamma_{\mu}
(3P^{\mu}+3Q^{\mu}+2L^{\mu})v(\boldsymbol{p}_l)\,.
$$

\textbf{diagram 2. (b)}
\begin{eqnarray}
{\cal A}^{+-}
  &=& \frac{1}{12F_0}\,G_FV_{us}^*{\overline u}(\boldsymbol{p}_{\nu})
        (1+\gamma^5)\gamma_{\mu}\,\frac{1}{s_l-M_{K^{\pm}}^2}\times \nonumber \\
  & & [\,3(s_{\pi}+t_{\pi}-u_{\pi})-(s_l-M_{K^{\pm}}^2 )
        +24Z_0e^2F_0^2\,]\,L^{\mu}v(\boldsymbol{p}_l)\,.
\end{eqnarray}

\textbf{diagram 2. (c)}
$$
{\cal A}^{+-}\,=\,-\frac{1}{F_0}\,G_FV_{us}^*{\overline
u}(\boldsymbol{p}_{\nu})(1+\gamma^5)\gamma_{\mu}\,
\frac{e^2F_0^2}{s_{\pi}}\,Q^{\mu}v(\boldsymbol{p}_l)\,.
$$

\textbf{diagram 2. (d)}
$$
{\cal A}^{+-}\,=\,-\frac{1}{F_0}\,G_FV_{us}^*{\overline
u}(\boldsymbol{p}_{\nu})(1+\gamma^5)\gamma_{\mu}\,\frac{t_{\pi}-u_{\pi}}{s_l-M_{K^{\pm}}^2}\,
\frac{e^2F_0^2}{s_{\pi}}\,L^{\mu}v(\boldsymbol{p}_l)\,.
$$

\textbf{diagram 2. (e)}
\begin{eqnarray}
{\cal A}^{+-}
&=& \frac{1}{F_0}\,G_FV_{us}^*{\overline u}(\boldsymbol{p}_{\nu})
        (1+\gamma^5)\,\frac{e^2F_0^2}{s_{\pi}}\,\frac{1}{M_{K^{\pm}}^2-m_l^2-2p\cdot p_{\nu}}\,\times
\nonumber \\ 
&& \left [(M_{K^{\pm}}^2-m_l^2-2p\cdot p_{\nu})\gamma_{\mu}Q^{\mu}\right. 
\nonumber \\ 
&+& \left.(Q\cdot L+Q\cdot N)\gamma_{\mu}L^{\mu}
        -2im_l\sigma_{\mu\nu}p_1^{\mu}p_2^{\nu}\right ]v(\boldsymbol{p}_l)\,. 
\nonumber
\end{eqnarray}

Note the cancellation between the $Q^{\mu}$ contributions from
diagrams \textbf{2. (c)} and \textbf{2. (e)}. Adding all this
together, the corrected form factors in the charged channel read,
\begin{eqnarray}
f^{+-}
  &=& F^{+-}\,, \nonumber \\
g^{+-}
  &=& G^{+-}\,, \nonumber \\
r^{+-}
  &=& R^{+-}+4Z_0e^2F_0^2\,\frac{M_{K^{\pm}}}{\sqrt{2}F_0}\,\frac{1}{s_l-M_{K^{\pm}}^2} \nonumber \\
  &-& \frac{2e^2F_0^2}{s_{\pi}}\left
  (\frac{t_{\pi}-u_{\pi}}{s_l-M_{K^{\pm}}^2}-
        \frac{Q\cdot L+Q\cdot N}{M_{K^{\pm}}^2-m_l^2-2p\cdot p_{\nu}}\right )\frac{M_{K^{\pm}}}{\sqrt{2}F_0}\,, \nonumber \\
T
  &=& \frac{4e^2F_0^2}{s_{\pi}}\,\frac{m_lM_{K^{\pm}}}{M_{K^{\pm}}^2-m_l^2-2p\cdot p_{\nu}}\,\frac{M_{K^{\pm}}}{\sqrt{2}F_0}\,, \nonumber
\end{eqnarray} 
with the charged form factors,
\begin{eqnarray}
F^{+-}
  &=& G^{+-} \nonumber \\
  &=& \frac{M_{K^{\pm}}}{\sqrt{2}F_0}\,, \nonumber \\
R^{+-}
  &=& \frac{M_{K^{\pm}}}{2\sqrt{2}F_0}\left
  (1+\frac{s_{\pi}+t_{\pi}-u_{\pi}}{s_l-M_{K^{\pm}}^2}\right )\,, \nonumber \\
H^{+-}
  &=& 0\,. \nonumber
\end{eqnarray}

The corrected form factors in the neutral channel are given by,
\begin{eqnarray} f^{00}
  &=& F^{00}-\frac{6\epsilon}{\sqrt{3}}\,\frac{M_{K^{\pm}}}{\sqrt{2}F_0}\,, \nonumber \\
g^{00}
  &=& G^{00}\,, \nonumber \\
r^{00}
  &=& R^{00}-\frac{\epsilon}{\sqrt{3}}\,\frac{M_{K^{\pm}}}{\sqrt{2}F_0}\times \nonumber \\
  & & \left [1+\frac{1}{s_l-M_{K^{\pm}}^2}\,(3s_{\pi}+2s_l-6M_K^2 )\right ]\,, \nonumber
\end{eqnarray} with the neutral form factors, \begin{eqnarray} F^{00}
  &=& -\frac{M_{K^{\pm}}}{\sqrt{2}F_0}\,, \nonumber \\
G^{00}
  &=& 0\,, \nonumber \\
R^{00}
  &=& -\frac{M_{K^{\pm}}}{2\sqrt{2}F_0}\left
  (1+\frac{s_{\pi}}{s_l-M_{K^{\pm}}^2}\right )\,, \nonumber \\
H^{00}
  &=& 0\,. \nonumber
\end{eqnarray}

The ones relative to the mixed channel have the following form,
\begin{eqnarray} f^{0-}
  &=& F^{0-}-\frac{3\epsilon}{\sqrt{3}}\,\frac{M_{K^{\pm}}}{F_0}\,, \nonumber \\
g^{0-}
  &=& G^{0-}\,, \nonumber \\
r^{0-}
  &=& R^{0-}+\frac{\Delta_{\pi}}{s_l-M_{K^{\pm}}^2}\,\frac{M_{K^{\pm}}}{2F_0} \nonumber \\
  &-& \frac{\epsilon}{\sqrt{3}}\,\frac{M_{K^{\pm}}}{2F_0}
        \left [1+\frac{1}{s_l-M_{K^{\pm}}^2}\,(3s_{\pi}+2s_l-6M_K^2 )\right ]\,, \nonumber
\end{eqnarray} with the mixed form factors, \begin{eqnarray} F^{0-}
  &=& 0\,, \nonumber \\
G^{0-}
  &=& \frac{M_{K^{\pm}}}{F_0}\,, \nonumber \\
R^{0-}
  &=& \frac{M_{K^{\pm}}}{2F_0}\, \frac{t_{\pi}-u_{\pi}}{s_l-M_{K^{\pm}}^2}\,, \nonumber \\
H^{0-}
  &=& 0\,. \nonumber
\end{eqnarray}

\section{Mixed channel at next-to-leading order}
\label{sec:next_to_leading} 

\begin{figure}[p]
\epsfxsize14cm \centerline{\epsffile{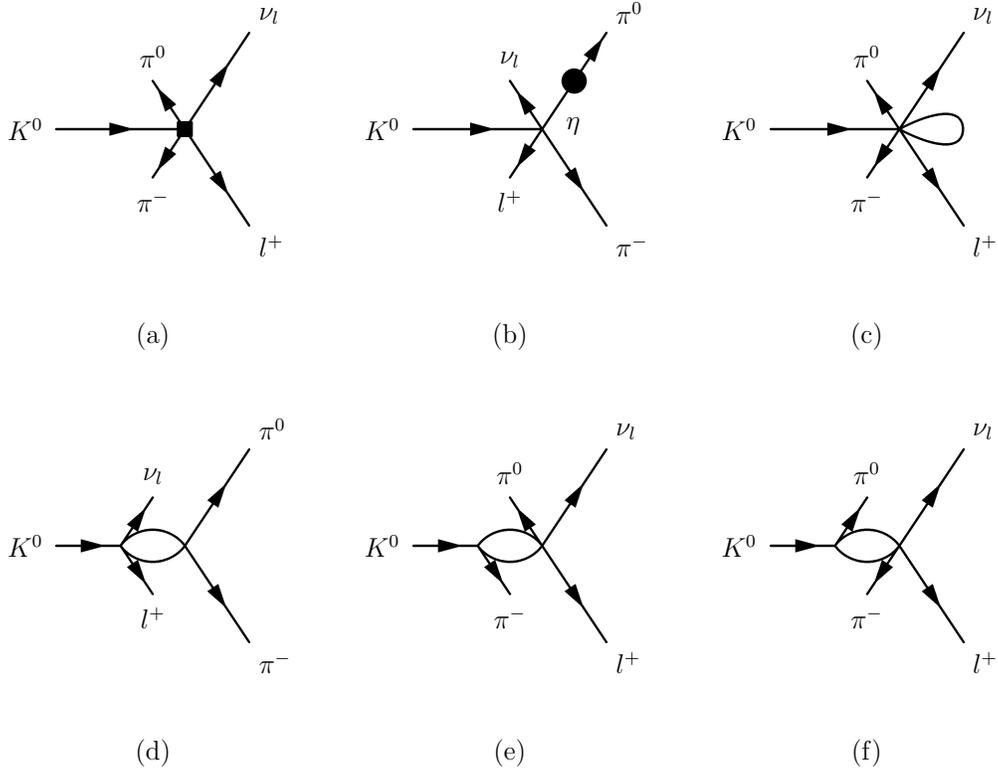}}
\caption{\label{fig:strong} Feynman diagrams representing the $K_{\ell 4}$ decay amplitude of the neutral kaon at one-loop. Is shown only the non photonic topology. Diagram (a) represents Born and counter-terms contributions. Diagram (b) accounts for $\pi^0-\eta$ mixing. Tadpole contribution is given by diagram (c). Diagrams (d), (e) and (f) stand for contributions from the $s$-, $t$- and $u$-channels, respectively.}
\end{figure}

We present here a one-loop calculation of the $K_{\ell 4}$ decay amplitude for the mixed channel including isospin breaking terms. Feynman diagrams representing the amplitude will be separated into two sets: photonic and non photonic diagrams. The non photonic set is drawn in figure~\ref{fig:strong}. 

The calculation of these diagrams is standard in field theory. The starting point is Lagrangian (\ref{Leff}). For the non linear realization of chiral symmetry we will use the exponential parametrization,
\begin{equation}
\label{eq:exponential_parametrization} 
u_R\,=\,u_L^{\dagger}\,=\,\mathrm{exp}\left\lbrace \dfrac{i\Phi}{2F_0}\right\rbrace \,,
\end{equation}  
where $\Phi$ is the linear realization of $SU(3)$ and can be decomposed in the basis of Gellmann-Low matrices $\lambda_a$ as,
$$
\Phi\,=\,\Phi^{\dagger}\,=\,\sum_{a=1}^8\lambda_a\phi^a\,.
$$
In terms of physical fields the matrix $\Phi$ can be written,
\begin{eqnarray}
\Phi_{11}
&=& \left( 1+\dfrac{\tilde{\epsilon}_2}{\sqrt{3}}\right) \pi^0+\left( -\tilde{\epsilon}_1+\dfrac{1}{\sqrt{3}}\right) \eta\,,
\nonumber \\
\Phi_{12}
&=& -\sqrt{2}\pi^+\,, \; \Phi_{13}\,=\,-\sqrt{2}K^+\,, \; \Phi_{21}\,=\,\sqrt{2}\pi^-\,, 
\nonumber \\
\Phi_{22}
&=& \left( -1+\dfrac{\tilde{\epsilon}_2}{\sqrt{3}}\right) \pi^0+\left( \tilde{\epsilon}_1+\dfrac{1}{\sqrt{3}}\right) \eta\,,
\nonumber \\
\Phi_{23}
&=& -\sqrt{2}K^0\,, \; \Phi_{31}\,=\,\sqrt{2}K^-\,, \; \Phi_{32}\,=\,-\sqrt{2}\overline{K}^0\,, 
\nonumber \\ 
\Phi_{33}
&=& -\dfrac{2}{\sqrt{3}}\left(\tilde{\epsilon}_2\pi^0+\eta\right) \,. 
\label{eq:linear_parametrization} 
\end{eqnarray}
The two mixing angles $\tilde{\epsilon}_1$ and $\tilde{\epsilon}_2$ relate $\phi^3$ and $\phi^8$ to the mass eigenstates $\pi^0$ and $\eta$, 
\begin{equation}
\left( 
\begin{array}{c}
\pi^0 \\
\eta
\end{array}
\right) \,\doteq\,
\left( 
\begin{array}{cc}
1                   & \tilde{\epsilon}_1 \\
-\tilde{\epsilon}_2 & 1 
\end{array}
\right) 
\left( 
\begin{array}{c}
\phi^3 \\
\phi^8
\end{array}
\right) \,.
\end{equation}
Notice that $\tilde{\epsilon}_1=\tilde{\epsilon}_2=\epsilon$ at leading order. The next step consists on expanding Lagrangian (\ref{Leff}) to fifth order in pseudoscalar fields, generating Feynman rules and drawing allowed topologies. We then calculate pseudoscalar propagators and derive masses and wave function renormalization constants. Finally, we expand the next-to-leading order Lagrangian to third order in pseudoscalar fields and obtain the counterterm contribution.  
 
Let us denote by $\delta F$ and $\delta G$ the next-to-leading
order corrections to the $F^{0-}$ and $G^{0-}$ form factors,
respectively, 
\begin{eqnarray} 
f^{0-}
&=& \frac{M_{K^{\pm}}}{F_0}\,\bigg (\,0+\delta F\,\bigg )\,,
\nonumber \\
g^{0-}
&=& \frac{M_{K^{\pm}}}{F_0}\,\bigg (\,1+\delta G\,\bigg )\,.
\nonumber
\end{eqnarray}
The expressions for $\delta F$ and $\delta G$ are lengthy. Therefore we will separate them to different contributions depending on the topology of the Feynman diagram representing a given contribution. 

\subsection{Born contribution}

This contribution is obtained from diagram (a) in figure~\ref{fig:strong}. We take the corresponding vertex from Lagrangian (\ref{Leff}) and multiply by the wave function renormalization constant factor to obtain,
\begin{eqnarray} 
\delta F
&=& -\frac{3\epsilon}{\sqrt{3}}\left\{1-\frac{1}{24F_0^2}\left
  [19A_0(M_{\pi})+3A_0(M_{\eta})+14A_0(M_K)\right ]\right. 
\nonumber \\
&-& \left.\frac{4}{F_0^2}\left [3(M_{\pi}^2 +2M_K^2 )L_4+(M_K^2 +2M_{\pi}^2
  )L_5\right ]\right\}\,, 
\nonumber \\
\delta G
&=& -\frac{1}{24F_0^2}\left [\left
  (5-\frac{6\epsilon}{\sqrt{3}}\right )A_0(M_{\pi^0})+3\left
  (1+\frac{2\epsilon}{\sqrt{3}}\right )A_0(M_{\eta})\right. 
\nonumber \\
&+& \left.14A_0(M_{\pi^{\pm}})+8A_0(M_{K^0})+6A_0(M_{K^{\pm}})\right ] \nonumber \\
&-& \frac{4}{F_0^2}\left [3(M_{\pi}^2 +2M_K^2 )L_4+(M_{K^0}^2 +2M_{\pi^0}^2
  )L_5\right ] 
\nonumber \\
&+& \frac{e^2}{2}\left [\frac{2}{M_{\pi}^2}\,A_0(M_{\pi})-\frac{1}{m_l^2}\,A_0(m_l)-\frac{1}{16\pi^2}\left
  (5+2\ln\frac{m_{\gamma}^2}{M_{\pi}^2}+2\ln\frac{m_{\gamma}^2}{m_l^2}\right
  )\right ] 
\nonumber \\
&-& \frac{e^2}{6}\left
  (24K_1+24K_2-12K_3+6K_4+16K_5+16K_6+3X_6\right )\,. 
\nonumber
\end{eqnarray}
The Born contribution is infrared divergent. This divergence emerges from the wave function renormalization constants of charged pion and lepton. 

\subsection{$\pi^0$ - $\eta$ mixing contribution}

Isospin breaking induces $\pi^0-\eta$ mixing by assigning non vanishing values to the off-diagonal matrix elements,
\begin{equation}
\langle 0|A_{\mu}^8|\pi^0(p)\rangle\,\doteq\,ip_{\mu}F_{\pi}\epsilon_1\,, \quad 
\langle 0|A_{\mu}^3|\eta (p)\rangle\,\doteq\,-ip_{\mu}F_{\eta}\epsilon_2\,.
\end{equation} 
The calculation of these matrix elements is staightforward and one obtains for the two mixing angles the following next-to-leading order expressions,
\begin{eqnarray}
\epsilon_1
&=& \epsilon_2+2\epsilon (M_K^2-M_{\pi}^2)C_1
\nonumber \\ 
&-& \dfrac{1}{3\sqrt{3}}\,\dfrac{e^2}{(4\pi )^2}\left[ 9Z_0\left( 1+\ln\dfrac{M_K^2}{\mu^2}\right) \right. 
\nonumber \\ 
&+& \left.2(4\pi )^2(6K_3^r-3K_4^r-2K_5^r-2K_6^r)\right] \,,
\nonumber \\ 
\epsilon_2
&=& \epsilon\left[ 1-3\mu_{\pi}+2\mu_K+\mu_{\eta}\right.
\nonumber \\ 
&+& \left.M_{\pi}^2C_1-\dfrac{32}{F_0^2}\,(M_K^2-M_{\pi}^2)(3L_7+L_8^r)\right] 
\nonumber \\ 
&-& \dfrac{2}{3\sqrt{3}}\,\dfrac{e^2}{(4\pi )^2}\,\dfrac{M_{\pi}^2}{M_{\eta}^2-M_{\pi}^2}\left[ 3Z_0\left( 1+\ln\dfrac{M_K^2}{\mu^2}\right) \right.
\nonumber \\ 
&+& \left.(4\pi )^2(6K_3^r-3K_4^r-2K_5^r-2K_6^r+2K_9^r+2K_{10}^r)\right]\,. 
\end{eqnarray} 
Herein, the quantity $C_1$ is defined by,
$$
C_1\,\doteq\,\dfrac{1}{16\pi^2F_0^2}\left( 1-\dfrac{M_{\pi}^2}{M_K^2-M_{\pi}^2}\,\ln\dfrac{M_K^2}{M_{\pi}^2}\right) \,,
$$
and the tadpole integrals are denoted by,
$$
\mu_P\,\doteq\,\dfrac{1}{32\pi^2F_0^2}\,\ln\dfrac{M_P^2}{\mu^2}\,.
$$
Let us relate the two sets of mixing angles, $(\epsilon_1,\epsilon_2)$ and $(\tilde{\epsilon}_1,\tilde{\epsilon}_2)$. To this end, one calculates the pseudoscalar mass matrix to next-to-leading order using the parametrization (\ref{eq:exponential_parametrization}) and (\ref{eq:linear_parametrization}). Then the diagonalization condition leads to,
\begin{eqnarray}
\tilde{\epsilon}_1
&=& \epsilon_1-\dfrac{2\epsilon}{3}\,(M_K^2-M_{\pi}^2)C_1
\nonumber \\ 
&+& \dfrac{Z_0}{\sqrt{3}}\,\dfrac{e^2}{(4\pi )^2}\left( 1+\ln\dfrac{M_K^2}{\mu^2}\right) +\dfrac{2}{\sqrt{3}}\,Z_0e^2\overline{\lambda}\,, 
\nonumber \\ 
\tilde{\epsilon}_2
&=& \epsilon_2+\dfrac{2\epsilon}{3}\,(M_K^2-M_{\pi}^2)C_1
\nonumber \\ 
&-& \dfrac{Z_0}{\sqrt{3}}\,\dfrac{e^2}{(4\pi )^2}\left( 1+\ln\dfrac{M_K^2}{\mu^2}\right) -\dfrac{2}{\sqrt{3}}\,Z_0e^2\overline{\lambda}\,. 
\nonumber 
\end{eqnarray} 
Since $\epsilon$ represents the $\pi^0-\eta$ mixing angle at leading chiral order it is interesting to ``renormalize'' $\epsilon$ by taking the mean value of the mixing angles and expand it to next-to-leading chiral order,
\begin{eqnarray}
\overline{\epsilon}
&\doteq & \frac{1}{2}\,(\tilde{\epsilon}_1+\tilde{\epsilon}_2)\,=\,
\frac{1}{2}\,(\epsilon_1+\epsilon_2)
\nonumber \\ 
&=& \epsilon^{(2)}+\epsilon_{\mathrm{str.}}^{(4)}
+\epsilon_{\mathrm{em.}}^{(4)}+\mathcal{O}(p^6)\,,
\end{eqnarray}
where,
\begin{eqnarray}
\epsilon^{(2)}
&=& \epsilon\,, 
\nonumber \\
\epsilon_{\mathrm{str.}}^{(4)}
&=& \epsilon\left[ -3\mu_{\pi}+2\mu_K+\mu_{\eta}\right.
\nonumber \\ 
&+& \left.M_K^2C_1-\dfrac{32}{F_0^2}\,(M_K^2-M_{\pi}^2)(3L_7+L_8^r)\right]\,, 
\nonumber \\
\epsilon_{\mathrm{em.}}^{(4)} 
&=& -\dfrac{2}{9\sqrt{3}}\,\dfrac{e^2}{(4\pi )^2}\,\dfrac{M_K^2}{M_{\eta}^2-M_{\pi}^2}\left[ 9Z_0\left( 1+\ln\dfrac{M_K^2}{\mu^2}\right) \right.
\nonumber \\ 
&+& \left.2(4\pi )^2(6K_3^r-3K_4^r-2K_5^r-2K_6^r)\right]
\nonumber \\
&-& \dfrac{2e^2}{9\sqrt{3}}\,\dfrac{M_{\pi}^2}{M_{\eta}^2-M_{\pi}^2}\,\times
\nonumber \\ 
&& (6K_3^r-3K_4^r-2K_5^r-2K_6^r+6K_9^r+6K_{10}^r)\,. 
\end{eqnarray}
Note that the preceding equations were derived in~\cite{Cirigliano:2001mk} neglecting terms proportional to $e^2M_{\pi}^2$.  
From the foregoing, one can treat $\pi^0-\eta$ mixing following two methods. The first consists on working with mass eigenstates, that is, with pseudoscalars diagonalizing the mass matrix~\cite{Ecker:1999kr}. In practice, this is reached by keeping $\tilde{\epsilon}_1$ and $\tilde{\epsilon}_2$ when deriving Feynman rules from Lagrangian (\ref{Leff}) with parametrization (\ref{eq:exponential_parametrization}) and (\ref{eq:linear_parametrization}). At the end one replaces $\tilde{\epsilon}_1$ and $\tilde{\epsilon}_2$ with their expressions quoted before. Notice that $\tilde{\epsilon}_1$ and $\tilde{\epsilon}_2$ are divergent quantities and, thanks to this divergence, the final result is finite. We will not follow the method just described and opt for the following one. Let us set $\tilde{\epsilon}_1$ and $\tilde{\epsilon}_2$ in parametrization (\ref{eq:exponential_parametrization}) and (\ref{eq:linear_parametrization}) to their leading order value, $\epsilon$. As usual, derive Feynman rules from Lagrangian (\ref{Leff}) and calculate one-particle-irreducible diagrams. After taking into account all contributions especially those coming from counter-terms, the final result still divergent naturally. This divergence can be cancelled by the one generated from diagram (b) in figure~\ref{fig:strong} which accounts for $\pi^0-\eta$ mixing. The contribution of this diagram to the $F$ and $G$ form factors is,  
\begin{eqnarray} 
\delta F
&=& \frac{1}{3F_0^2}\,\frac{1}{M_{\pi^0}^2-M_{\eta}^2}\times 
\nonumber \\ 
&& \left\{\frac{2\epsilon}{\sqrt{3}}\left (5M_K^2 -11M_{\pi}^2
  \right )A_0(M_{\pi^0})-\frac{6\epsilon}{\sqrt{3}}\left (M_K^2 -M_{\pi}^2
  \right )A_0(M_{\eta})\right. 
\nonumber \\
&-& \left [2M_{K^0}^2 +M_{\pi^0}^2 -\frac{2\epsilon}{\sqrt{3}}\left (M_K^2
  +2M_{\pi}^2 \right )\right ]A_0(M_{K^0}) 
\nonumber \\
&+& \left [2M_{K^0}^2 +M_{\pi^0}^2 -\frac{6\epsilon}{\sqrt{3}}\left (M_K^2
  -2M_{\pi}^2 \right )\right ]A_0(M_{K^{\pm}}) 
\nonumber \\
&-& \frac{384\epsilon}{\sqrt{3}}\left (M_K^2 -M_{\pi}^2 \right )^2\left
  (3L_7+L_8\right ) 
\nonumber \\
&-& \left.2e^2F_0^2M_{\pi}^2 \left
  (6K_3-3K_4-2K_5-2K_6+2K_9+2K_{10}\right )\right\}\,, 
\nonumber \\
\delta G
&=& 0\,.
\end{eqnarray} 
We have checked that the two methods are completly equivalent.

\subsection{Counter-terms contribution}

The contribution in question follows from diagram (a) in figure~\ref{fig:strong} with the help of the next-to-leading order Lagrangian and reads,
\begin{eqnarray} 
\delta F
  &=& \frac{4}{F_0^2}\left [p\cdot p_1-p\cdot
  p_2+\frac{\epsilon}{\sqrt{3}}\left (p\cdot p_1+p\cdot
  p_2-4p_1\cdot p_2\right )\right ]L_3 \nonumber \\
  &-& \frac{6}{F_0^2}\,\frac{\epsilon}{\sqrt{3}}\left [4(M_{\pi}^2
  +2M_K^2 )L_4+2(M_K^2 +2M_{\pi}^2 )L_5+s_lL_9\right ] \nonumber \\
  &+& \frac{e^2}{3}\left (-6K_3+3K_4+2K_5+2K_6-6X_1\right )\,,
  \nonumber \\ 
\delta G
  &=& -\frac{4}{F_0^2}\left [p\cdot p_1+p\cdot
  p_2+\frac{\epsilon}{\sqrt{3}}\left (p\cdot p_1-p\cdot
  p_2\right )\right ]L_3 \nonumber \\
  &+& \frac{2}{F_0^2}\left [4(M_{\pi}^2
  +2M_K^2 )L_4+2(M_{K^0}^2 +2M_{\pi^0}^2 )L_5+s_lL_9\right ] \nonumber \\
  &+& \frac{e^2}{9}\left (24K_1+24K_2-18K_3+9K_4+20K_5+20K_6+18K_{12}-6X_1\right
  )\,.
  \nonumber
\end{eqnarray}

\subsection{Tadpole contribution}

Tadpoles are shawn in diagram (c) of figure~\ref{fig:strong} and contribute to the form factors by the following,
\begin{eqnarray} 
\delta F
&=& -\frac{1}{4F_0^2}\left [\left (1+\frac{5\epsilon}{\sqrt{3}}\right )A_0(M_{\pi^0})
   +\frac{6\epsilon}{\sqrt{3}}\,A_0(M_{\eta})\right. 
\nonumber \\
&+& \left.\frac{8\epsilon}{\sqrt{3}}\,A_0(M_{K^0})-\left (1-\frac{5\epsilon}{\sqrt{3}}\right )A_0(M_{\pi^{\pm}})+\frac{6\epsilon}{\sqrt{3}}\,A_0(M_{K^{\pm}})\right ]\,, 
\nonumber \\  
\delta G
&=& \frac{1}{4F_0^2}\left [2\left (1+\frac{2\epsilon}{\sqrt{3}}\right )A_0(M_{\pi^0})
   +\left (1-\frac{3\epsilon}{\sqrt{3}}\right )A_0(M_{\eta})\right. 
\nonumber \\
&+& \left.2\left (1-\frac{\epsilon}{\sqrt{3}}\right )A_0(M_{K^0})+\left
   (3+\frac{\epsilon}{\sqrt{3}}\right )A_0(M_{\pi^{\pm}})+2A_0(M_{K^{\pm}})\right ]\,. 
\nonumber
\end{eqnarray}

\subsection{The $s$-channel contribution}

The remaining contribution to the non photonic part of the decay amplitude comes from loop diagrams with two pseudoscalar propagators. This two-point function contribution will be separated in three parts depending on the Lorentz scalar governing its underlying kinematics. The $s$-channel contribution comes from diagram (d) in figure~\ref{fig:strong} and reads,
\begin{eqnarray} 
\delta F
&=& -\frac{1}{3F_0^2}\left\{-A_0(M_{\pi^{\pm}})
-\frac{3\epsilon}{\sqrt{3}}\,A_0(M_K)\right. 
\nonumber \\
&+& \left [M_{\pi^0}^2 -2p_1\cdot p_2+\frac{3\epsilon}{\sqrt{3}}\left
  (M_{\pi}^2 -2p_1\cdot p_2\right )\right ]B_0(-p_1-p_2,M_{\pi^0},M_{\pi^{\pm}}) 
\nonumber \\
&-& \frac{6\epsilon}{\sqrt{3}}\left (M_{\eta}^2 -M_{\pi}^2 -p_1\cdot
  p_2\right )B_0(-p_1-p_2,M_{\eta},M_{\pi}) 
\nonumber \\
&+& 2\left [3M_{\pi^0}^2 -M_{\pi^{\pm}}^2 -p_1\cdot p_2\right. 
\nonumber \\ 
&+& \left.\frac{3\epsilon}{\sqrt{3}}\left (M_{\pi}^2 +p_1\cdot p_2\right
  )\right ]B_1(-p_1-p_2,M_{\pi^0},M_{\pi^{\pm}}) 
\nonumber \\
&-& \frac{6\epsilon}{\sqrt{3}}\left (M_{\pi}^2 +p_1\cdot p_2\right )B_1(-p_1-p_2,M_{\eta},M_{\pi}) 
\nonumber \\
&+& \frac{6\epsilon}{\sqrt{3}}\left (M_K^2 -M_{\pi}^2 -p_1\cdot
  p_2\right )B_1(-p_1-p_2,M_K,M_K) 
\nonumber \\
&+& 2B_{00}(-p_1-p_2,M_{\pi^0},M_{\pi^{\pm}})
+\frac{3\epsilon}{\sqrt{3}}\,B_{00}(-p_1-p_2,M_K,M_K) 
\nonumber \\
&+& 4(2M_{\pi^0}^2 -M_{\pi^{\pm}}^2 +p_1\cdot p_2)B_{11}(-p_1-p_2,M_{\pi^0},M_{\pi^{\pm}}) 
\nonumber \\
&-& \left.3\left [\Delta_{\pi}-\frac{2\epsilon}{\sqrt{3}}
  \left (M_{\pi}^2 +p_1\cdot p_2\right )\right ]
  B_{11}(-p_1-p_2,M_K,M_K)\right\}\,, 
\nonumber \\ 
\delta G
&=& -\frac{1}{F_0^2}\left [2B_{00}(-p_1-p_2,M_{\pi^0},M_{\pi^{\pm}})
+B_{00}(-p_1-p_2,M_{K^0},M_{K^{\pm}})\right ]\,. 
\nonumber
\end{eqnarray}

\subsection{The $t$-channel contribution}

Diagram (f) in figure~\ref{fig:strong} generates the somewhat lengthy $t$-channel contribution,
\begin{eqnarray} 
\delta F
&=& -\frac{1}{12F_0^2}\left\{\left
  (1-\frac{3\epsilon}{\sqrt{3}}\right )A_0(M_{K^0})-3\left
  (1+\frac{\epsilon}{\sqrt{3}}\right )A_0(M_{K^{\pm}})\right. 
\nonumber \\
&+& \left [2M_{\pi^0}^2 +p\cdot p_1\right. 
\nonumber \\ 
&+& \left.\frac{3\epsilon}{\sqrt{3}}\left
  (4M_K^2 -2M_{\pi}^2 -p\cdot p_1\right )\right ]B_0(p_1-p,M_{\pi^0},M_{K^0}) 
\nonumber \\
&+& \frac{3}{2}\left [M_{\eta}^2 -M_{\pi^0}^2 -2p\cdot p_1\right. 
\nonumber \\ 
&+& \left.\frac{3\epsilon}{\sqrt{3}}\left (M_{\eta}^2 -M_{\pi}^2 +2p\cdot
  p_1\right )\right ]B_0(p_1-p,M_{\eta},M_{K^0}) 
\nonumber \\
&-& \left [M_{K^0}^2 +2p\cdot p_1\right. 
\nonumber \\ 
&+& \left.\frac{3\epsilon}{\sqrt{3}}\left
  (3M_K^2 -4M_{\pi}^2 -4p\cdot p_1\right )\right ]B_1(p_1-p,M_{\pi^0},M_{K^0}) 
\nonumber \\
&+& \frac{3}{2}\left [M_{\eta}^2 +2M_{K^0}^2 -5M_{\pi^0}^2\right.
\nonumber \\ 
&+& \left.\frac{2\epsilon}{\sqrt{3}}\left (M_{\eta}^2 -M_K^2 +3M_{\pi}^2 -2p\cdot
  p_1\right )\right ]B_1(p_1-p,M_{\eta},M_{K^0}) 
\nonumber \\
&+& 6\left [p\cdot p_1-\frac{\epsilon}{\sqrt{3}}\left (2M_K^2
  -2M_{\pi}^2 -p\cdot p_1\right )\right ]B_1(p_1-p,M_{\pi^{\pm}},M_{K^{\pm}}) 
\nonumber \\
&+& 4\left (1-\frac{9\epsilon}{\sqrt{3}}\right )B_{00}(p_1-p,M_{\pi^0},M_{K^0}) 
\nonumber \\
&+& 12\left (1-\frac{\epsilon}{\sqrt{3}}\right )B_{00}(p_1-p,M_{\eta},M_{K^0})
\nonumber \\ 
&+& 12\left (1-\frac{2\epsilon}{\sqrt{3}}\right
  )B_{00}(p_1-p,M_{\pi^{\pm}},M_{K^{\pm}}) 
\nonumber \\
&+& \left [M_{K^0}^2 -2M_{\pi^0}^2 +p\cdot p_1\right. 
\nonumber \\ 
&-& \left.\frac{9\epsilon}{\sqrt{3}}\left (M_K^2 -2M_{\pi}^2 +p\cdot
  p_1\right )\right ]B_{11}(p_1-p,M_{\pi^0},M_{K^0}) 
\nonumber \\
&+& 3\left [M_{K^0}^2 -2M_{\pi^0}^2 +p\cdot p_1\right. 
\nonumber \\
&-& \left.\frac{\epsilon}{\sqrt{3}}\left (M_K^2 -2M_{\pi}^2 +p\cdot p_1\right
  )\right ]B_{11}(p_1-p,M_{\eta},M_{K^0}) 
\nonumber \\
&-& 6\left [p\cdot p_1-M_{K^0}^2\right. 
\nonumber \\ 
&+& \left.\left.\frac{\epsilon}{\sqrt{3}}\left (M_K^2
  -2M_{\pi}^2 +p\cdot p_1\right )\right ]B_{11}(p_1-p,M_{\pi^{\pm}},M_{K^{\pm}})\right\}\,, 
\nonumber \\ 
\delta G
&=& -\frac{1}{12F_0^2}\left\{-\left
  (1-\frac{3\epsilon}{\sqrt{3}}\right )A_0(M_{K^0})+3\left
  (1+\frac{\epsilon}{\sqrt{3}}\right )A_0(M_{K^{\pm}})\right. 
\nonumber \\
&-& \left [2M_{\pi^0}^2 +p\cdot p_1\right. 
\nonumber \\ 
&+& \left.\frac{3\epsilon}{\sqrt{3}}\left
  (4M_K^2 -2M_{\pi}^2 -p\cdot p_1\right )\right ]B_0(p_1-p,M_{\pi^0},M_{K^0}) 
\nonumber \\
&-& \frac{3}{2}\left [M_{\eta}^2 -M_{\pi^0}^2 -2p\cdot
  p_1\right. 
\nonumber \\ 
&+& \left.\frac{3\epsilon}{\sqrt{3}}\left (M_{\eta}^2 -M_{\pi}^2 +2p\cdot
  p_1\right )\right ]B_0(p_1-p,M_{\eta},M_{K^0}) 
\nonumber \\
&+& \left [M_{K^0}^2 +2p\cdot p_1\right. 
\nonumber \\ 
&+& \left.\frac{3\epsilon}{\sqrt{3}}\left
  (3M_K^2 -4M_{\pi}^2 -4p\cdot p_1\right )\right ]B_1(p_1-p,M_{\pi^0},M_{K^0}) 
\nonumber \\
&-& \frac{3}{2}\left [M_{\eta}^2 +2M_{K^0}^2 -5M_{\pi^0}^2\right. 
\nonumber \\ 
&+& \left.\frac{2\epsilon}{\sqrt{3}}\left (M_{\eta}^2 -M_K^2 +3M_{\pi}^2 -2p\cdot
  p_1\right )\right ]B_1(p_1-p,M_{\eta},M_{K^0}) 
\nonumber \\
&-& 6\left [p\cdot p_1-\frac{\epsilon}{\sqrt{3}}\left (2M_K^2
  -2M_{\pi}^2 -p\cdot p_1\right )\right ]B_1(p_1-p,M_{\pi^{\pm}},M_{K^{\pm}}) 
\nonumber \\
&+& 2\left (1-\frac{9\epsilon}{\sqrt{3}}\right )B_{00}(p_1-p,M_{\pi^0},M_{K^0}) 
\nonumber \\
&+& 6\left (1-\frac{\epsilon}{\sqrt{3}}\right
  )B_{00}(p_1-p,M_{\eta},M_{K^0})
-\frac{12\epsilon}{\sqrt{3}}\,B_{00}(p_1-p,M_{\pi},M_K) 
\nonumber \\
&-& \left [M_{K^0}^2 -2M_{\pi^0}^2 +p\cdot
  p_1\right. 
\nonumber \\ 
&-& \left.\frac{9\epsilon}{\sqrt{3}}\left (M_K^2 -2M_{\pi}^2 +p\cdot
  p_1\right )\right ]B_{11}(p_1-p,M_{\pi^0},M_{K^0}) 
\nonumber \\
&-& 3\left [M_{K^0}^2 -2M_{\pi^0}^2 +p\cdot
  p_1\right. 
\nonumber \\ 
&-& \left.\frac{\epsilon}{\sqrt{3}}\left (M_K^2 -2M_{\pi}^2 +p\cdot p_1\right
  )\right ]B_{11}(p_1-p,M_{\eta},M_{K^0}) 
\nonumber \\
&+& 6\left [p\cdot p_1-M_{K^0}^2\right. 
\nonumber \\ 
&+& \left.\left.\frac{\epsilon}{\sqrt{3}}\left (M_K^2
  -2M_{\pi}^2 +p\cdot p_1\right )\right ]B_{11}(p_1-p,M_{\pi^{\pm}},M_{K^{\pm}})\right\}\,. 
\nonumber
\end{eqnarray}

\subsection{The $u$-channel contribution}

Finally, the $u$-channel contribution follows from diagram (e) in figure~\ref{fig:strong},
\begin{eqnarray} 
\delta F
&=& \frac{1}{12F_0^2}\left\{-2A_0(M_{K^0})\right. 
\nonumber \\
&+& 3\left [p\cdot p_2+\frac{\epsilon}{\sqrt{3}}\left
  (2M_K^2 -2M_{\pi}^2 +5p\cdot p_2\right )\right ]B_0(p_2-p,M_{\pi^0},M_{K^{\pm}}) 
\nonumber \\
&+& \frac{3}{2}\left [M_{\eta}^2 -M_{\pi^0}^2 -2p\cdot
  p_2\right. 
\nonumber \\ 
&+& \left.\frac{2\epsilon}{\sqrt{3}}\left (M_{\eta}^2 -M_{\pi}^2 -5p\cdot
  p_2\right )\right ]B_0(p_2-p,M_{\eta},M_{K^{\pm}}) 
\nonumber \\
&+& 2\left [M_{\pi^{\pm}}^2 -p\cdot p_2-\frac{3\epsilon}{\sqrt{3}}\left
  (M_{\pi}^2 -p\cdot p_2\right )\right ]B_0(p_2-p,M_{\pi^{\pm}},M_{K^0}) 
\nonumber \\
&+& 3\left [M_{K^0}^2 +\frac{\epsilon}{\sqrt{3}}\left
  (9M_K^2 -4M_{\pi}^2 \right )\right ]B_1(p_2-p,M_{\pi^0},M_{K^{\pm}}) 
\nonumber \\
&+& \frac{3}{2}\left [M_{\eta}^2 +2M_{K^0}^2 -4M_{\pi^{\pm}}^2 -M_{\pi^0}^2\right. 
\nonumber \\
&+& \left.\frac{2\epsilon}{\sqrt{3}}\left (M_{\eta}^2 -M_K^2 -5M_{\pi}^2 \right )\right ]B_1(p_2-p,M_{\eta},M_{K^{\pm}}) 
\nonumber \\
&-& 4\left [M_{K^0}^2 -p\cdot p_2-\frac{3\epsilon}{\sqrt{3}}\left (M_K^2
  -M_{\pi}^2 \right )\right ]B_1(p_2-p,M_{\pi^{\pm}},M_{K^0}) 
\nonumber \\
&+& 6\left (1+\frac{8\epsilon}{\sqrt{3}}\right )B_{00}(p_2-p,M_{\pi^0},M_{K^{\pm}}) 
\nonumber \\
&+& 6\left (2+\frac{\epsilon}{\sqrt{3}}\right
  )B_{00}(p_2-p,M_{\eta},M_{K^{\pm}})
\nonumber \\ 
&+& 10\left (1+\frac{3\epsilon}{\sqrt{3}}\right
  )B_{00}(p_2-p,M_{\pi^{\pm}},M_{K^0}) 
\nonumber \\
&-& 3\left [-M_{K^0}^2 +p\cdot p_2\right. 
\nonumber \\ 
&+& \left.\frac{\epsilon}{\sqrt{3}}\left (-7M_K^2 +2M_{\pi}^2 +5p\cdot
  p_2\right )\right ]B_{11}(p_2-p,M_{\pi^0},M_{K^{\pm}}) 
\nonumber \\
&-& 3\left [-M_{K^0}^2 +2M_{\pi^{\pm}}^2 -p\cdot p_2\right. 
\nonumber \\ 
&+& \left.\frac{\epsilon}{\sqrt{3}}\left (M_K^2 +4M_{\pi}^2 -5p\cdot p_2\right
  )\right ]B_{11}(p_2-p,M_{\eta},M_{K^{\pm}}) 
\nonumber \\
&-& 2\left [M_{\pi^{\pm}}^2 -2M_{K^0}^2 +p\cdot p_2\right. 
\nonumber \\ 
&+& \left.\left.\frac{3\epsilon}{\sqrt{3}}\left (-2M_K^2
  +M_{\pi}^2 +p\cdot p_2\right )\right ]B_{11}(p_2-p,M_{\pi^{\pm}},M_{K^0})\right\}\,, 
\nonumber \\ 
\delta G
&=& \frac{1}{12F_0^2}\left\{-2A_0(M_{K^0})\right. 
\nonumber \\
&+& 3\left [p\cdot p_2+\frac{\epsilon}{\sqrt{3}}\left
  (2M_K^2 -2M_{\pi}^2 +5p\cdot p_2\right )\right ]B_0(p_2-p,M_{\pi^0},M_{K^{\pm}}) 
\nonumber \\
&+& \frac{3}{2}\left [M_{\eta}^2 -M_{\pi^0}^2 -2p\cdot p_2\right. 
\nonumber \\ 
&+& \left.\frac{2\epsilon}{\sqrt{3}}\left (M_{\eta}^2 -M_{\pi}^2 -5p\cdot
  p_2\right )\right ]B_0(p_2-p,M_{\eta},M_{K^{\pm}}) 
\nonumber \\
&+& 2\left [M_{\pi^{\pm}}^2 -p\cdot p_2-\frac{3\epsilon}{\sqrt{3}}\left
  (M_{\pi}^2 -p\cdot p_2\right )\right ]B_0(p_2-p,M_{\pi^{\pm}},M_{K^0}) 
\nonumber \\
&+& 3\left [M_{K^0}^2 +\frac{\epsilon}{\sqrt{3}}\left
  (9M_K^2 -4M_{\pi}^2 \right )\right ]B_1(p_2-p,M_{\pi^0},M_{K^{\pm}}) 
\nonumber \\
&+& \frac{3}{2}\left [M_{\eta}^2 +2M_{K^0}^2 -4M_{\pi^{\pm}}^2 -M_{\pi^0}^2\right. 
\nonumber \\ 
&+& \left.\frac{2\epsilon}{\sqrt{3}}\left (M_{\eta}^2 -M_K^2 -5M_{\pi}^2 \right )\right ]B_1(p_2-p,M_{\eta},M_{K^{\pm}}) 
\nonumber \\
&-& 4\left [M_{K^0}^2 -p\cdot p_2-\frac{3\epsilon}{\sqrt{3}}\left (M_K^2
  -M_{\pi}^2 \right )\right ]B_1(p_2-p,M_{\pi^{\pm}},M_{K^0}) 
\nonumber \\
&-& \frac{6\epsilon}{\sqrt{3}}\,B_{00}(p_2-p,M_{\pi},M_K) 
\nonumber \\
&-& 6\left (1+\frac{2\epsilon}{\sqrt{3}}\right
  )B_{00}(p_2-p,M_{\eta},M_{K^{\pm}})
\nonumber \\ 
&-& 2\left (1+\frac{3\epsilon}{\sqrt{3}}\right
  )B_{00}(p_2-p,M_{\pi^{\pm}},M_{K^0}) 
\nonumber \\
&-& 3\left [-M_{K^0}^2 +p\cdot
  p_2\right. 
\nonumber \\ 
&+& \left.\frac{\epsilon}{\sqrt{3}}\left (-7M_K^2 +2M_{\pi}^2 +5p\cdot
  p_2\right )\right ]B_{11}(p_2-p,M_{\pi^0},M_{K^{\pm}}) 
\nonumber \\
&-& 3\left [-M_{K^0}^2 +2M_{\pi^{\pm}}^2 -p\cdot
  p_2\right. 
\nonumber \\ 
&+& \left.\frac{\epsilon}{\sqrt{3}}\left (M_K^2 +4M_{\pi}^2 -5p\cdot p_2\right
  )\right ]B_{11}(p_2-p,M_{\eta},M_{K^{\pm}}) 
\nonumber \\
&-& 2\left [M_{\pi^{\pm}}^2 -2M_{K^0}^2 +p\cdot p_2\right. 
\nonumber \\ 
&+& \left.\left.\frac{3\epsilon}{\sqrt{3}}\left (-2M_K^2
  +M_{\pi}^2 +p\cdot p_2\right )\right ]B_{11}(p_2-p,M_{\pi^{\pm}},M_{K^0})\right\}\,. 
\nonumber
\end{eqnarray}

\subsection{Soft virtual photon contribution}

\begin{figure}[p]
\epsfxsize14cm \centerline{\epsffile{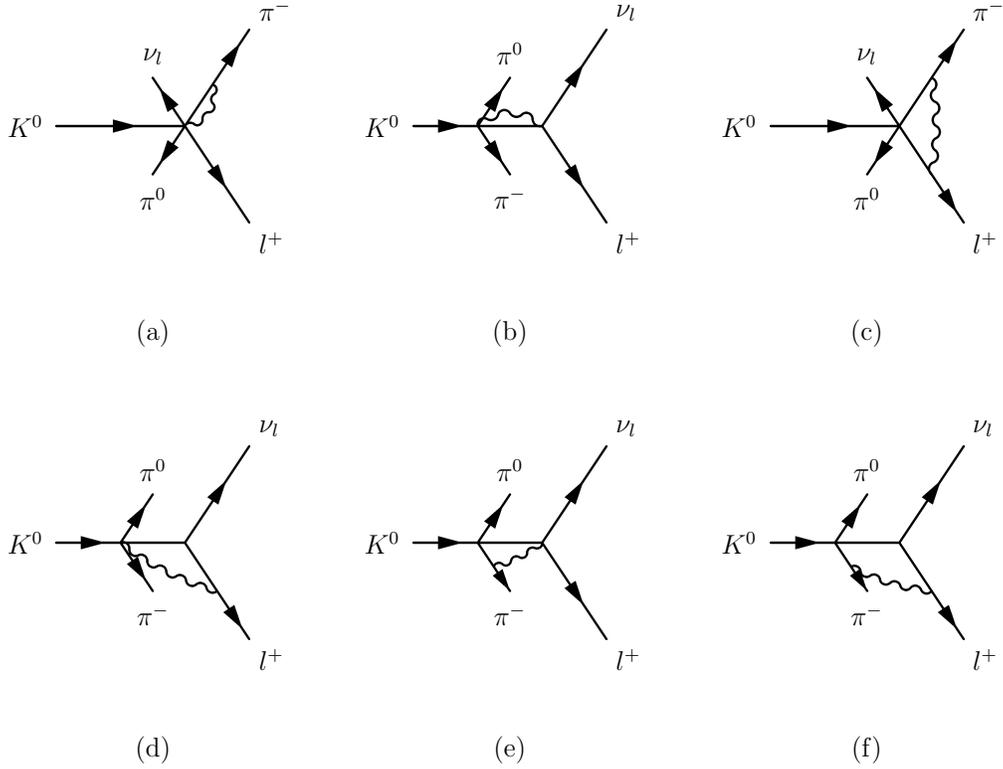}}
\caption{\label{fig:virtual} Virtual photons in Feynman diagrams for the $K_{\ell 4}$ decay in the mixed channel. Diagrams (a) and (b) represent tadpoles. Diagram (c) generates three-point functions of type leg-leg photon exchange. Diagrams (d) and (e) generate three-point functions of type vertex-leg. Four-point functions of type leg-leg are generated from diagram (f).}
\end{figure}

The various topologies of Feynman diagrams containing a virtual photon are drawn in figure~\ref{fig:virtual}.

Due to important cancellations between the different contributions from these diagrams we will present the result in a compact form,
\begin{eqnarray} 
\delta F
&=& \frac{e^2}{2}\left\{2B_0(-p_2,0,M_{\pi})-2B_0(p_2+p_l,m_l,M_{\pi})\right. \nonumber \\
&-& m_l^2C_1(-p_l,-p_l-p_{\nu},0,m_l,M_K)-m_l^2C_1(-p_l,p_2,0,m_l,M_{\pi}) 
\nonumber \\
&-& m_l^2C_2(-p_l,-p_l-p_{\nu},0,m_l,M_K)+4p_2\cdot p_lC_2(-p_l,p_2,0,m_l,M_{\pi}) \nonumber \\
&-& m_l^2(M_K^2 -s_l+2p\cdot p_1-2p\cdot p_2)D_1(-p_l,p_2,-p_l-p_{\nu},0,m_l,M_{\pi},M_K) 
\nonumber \\
&-& m_l^2(M_K^2 -s_l+2p\cdot p_1-2p\cdot
  p_2)D_3(-p_l,p_2,-p_l-p_{\nu},0,m_l,M_{\pi},M_K) 
\nonumber \\
&-& 2m_l^2p_1\cdot p_lD_{11}(-p_l,p_2,-p_l-p_{\nu},0,m_l,M_{\pi},M_K) 
\nonumber \\
&+& 2m_l^2p_1\cdot p_2D_{12}(-p_l,p_2,-p_l-p_{\nu},0,m_l,M_{\pi},M_K) 
\nonumber \\
&+& 2m_l^2(M_{\pi}^2 -p\cdot p_1+p_1\cdot p_2-p_1\cdot p_l)\times
\nonumber \\ 
&& \qquad D_{13}(-p_l,p_2,-p_l-p_{\nu},0,m_l,M_{\pi},M_K) 
\nonumber \\
&+& 2m_l^2p_1\cdot p_2D_{23}(-p_l,p_2,-p_l-p_{\nu},0,m_l,M_{\pi},M_K) 
\nonumber \\
&+& \left.2m_l^2(M_{\pi}^2 -p\cdot p_1+p_1\cdot p_2)D_{33}(-p_l,p_2,-p_l-p_{\nu},0,m_l,M_{\pi},M_K)\right\rbrace \,, 
\nonumber \\ 
\delta G
&=& -\frac{e^2}{2}\left\{-2B_0(-p_l,0,m_l)\right. 
\nonumber \\ 
&+& 8p_2\cdot p_lC_0(-p_l,p_2,m_{\gamma},m_l,M_{\pi})
+m_l^2C_1(-p_l,-p_l-p_{\nu},0,m_l,M_K)
\nonumber \\ 
&-& (m_l^2-4p_2\cdot p_l)C_1(-p_l,p_2,0,m_l,M_{\pi}) 
\nonumber \\
&+& m_l^2C_2(-p_l,-p_l-p_{\nu},0,m_l,M_K)+4p_2\cdot p_lC_2(-p_l,p_2,0,m_l,M_{\pi}) 
\nonumber \\
&-& m_l^2(M_K^2 -s_l+2p\cdot p_1-2p\cdot
  p_2)D_1(-p_l,p_2,-p_l-p_{\nu},0,m_l,M_{\pi},M_K) 
\nonumber \\
&-& m_l^2(M_K^2 -s_l+2p\cdot p_1-2p\cdot
  p_2)D_3(-p_l,p_2,-p_l-p_{\nu},0,m_l,M_{\pi},M_K) 
\nonumber \\
&-& 2m_l^2p_1\cdot p_lD_{11}(-p_l,p_2,-p_l-p_{\nu},0,m_l,M_{\pi},M_K)
\nonumber \\
&+& 2m_l^2p_1\cdot p_2D_{12}(-p_l,p_2,-p_l-p_{\nu},0,m_l,M_{\pi},M_K)
\nonumber \\
&+& 2m_l^2(M_{\pi}^2 -p\cdot p_1+p_1\cdot p_2-p_1\cdot p_l)\times
\nonumber \\ 
&& \qquad D_{13}(-p_l,p_2,-p_l-p_{\nu},0,m_l,M_{\pi},M_K) 
\nonumber \\
&+& 2m_l^2p_1\cdot p_2D_{23}(-p_l,p_2,-p_l-p_{\nu},0,m_l,M_{\pi},M_K)
\nonumber \\
&+& \left.2m_l^2(M_{\pi}^2 -p\cdot p_1+p_1\cdot p_2)D_{33}(-p_l,p_2,-p_l-p_{\nu},0,m_l,M_{\pi},M_K)\right\}\,. 
\nonumber
\end{eqnarray}
The infrared divergence in form factors is contained in loop functions with $m_{\gamma}$ in the argument.

\section{Soft photon bremsstrahlung}
\label{sec:soft_photon_bremsstrahlung} 

Virtual photon corrections to $K_{\ell 4}$ decay rate generate infrared
divergencies. These cancel, order by order in perturbation theory,
with the ones coming from real bremsstrahlung corrections.
Assume that the emitted photons are soft, that is, their energies
are smaller than any detector resolution, $\omega$. It follows
that radiative and non-radiative decays cannot be distinguished
experimentally and emission of real soft photons should be taken
into account. Note however that only single soft photon radiation
is needed to one-loop accuracy.

\subsection{The decay amplitude}
\label{sec:The decay amplitude}

\begin{figure}[p]
\epsfxsize14cm \centerline{\epsffile{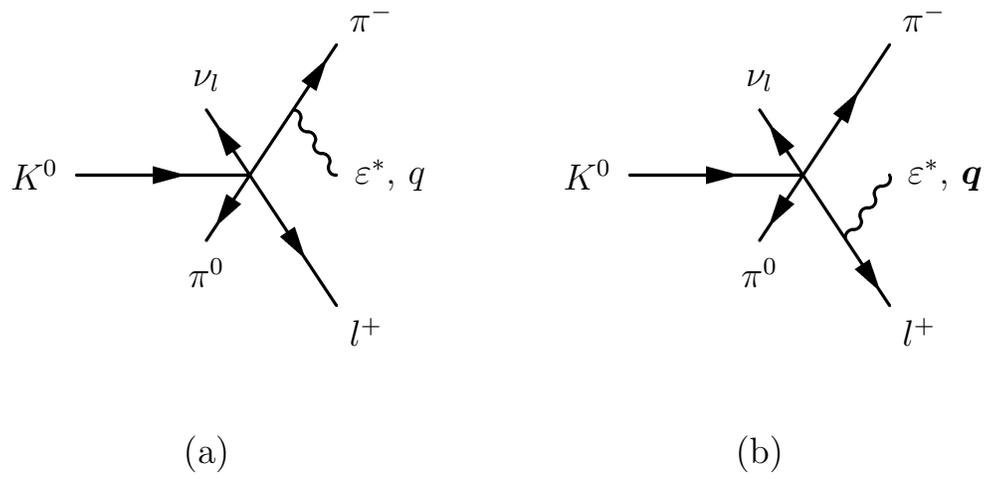}}
\caption{\label{fig:bremsstrahlung} Feynman diagrams representing the
contribution of $F$ and $G$ form factors to the bremsstrahlung
amplitude.}
\end{figure}

A general feature of photon bremsstrahlung is that, in the soft
photon approximation, the bremsstrahlung amplitude is proportional
to the Born amplitude. Since we deal only with isospin breaking
corrections to the $F$ and $G$ form factors, the Born amplitude is
taken, all along this section, to be,
$$
{\cal
A}_{\mathrm{B}}^{0-}\,=\,-\frac{1}{\sqrt{2}F_0}\,G_FV_{us}^*{\bar
u}(p_{\nu})\gamma_{\mu}(1-\gamma^5)v(p_l)Q^{\mu}\,.
$$
The contribution of form factors $F$ and $G$ to the Bremsstrahlung
amplitude can be read off from diagrams in Fig.~\ref{fig:bremsstrahlung}.

Let $\varepsilon$ and $q$ be, respectively, the polarization
vector and the momentum of the radiated photon. The evaluation of
diagrams in Fig.~\ref{fig:bremsstrahlung} is straightforward and read, to first
order in the photon energy, 
\begin{equation} 
\label{eq:bremsstrahlung amplitude} 
{\cal A}^{0-\gamma}\,=\,e{\cal
A}_{\mathrm{B}}^{0-}\left (\frac{p_l\cdot\varepsilon^*}{p_l\cdot
q}-\frac{p_2\cdot\varepsilon^*}{p_2\cdot
q}\right )+{\cal O}(q)\,. 
\end{equation}

Squaring the matrix element (\ref{eq:bremsstrahlung amplitude})
and summing over polarizations, we obtain 
\begin{eqnarray}
\sum_{\mathrm{pol.}}|{\cal A}^{0-\gamma}|^2
&=& -e^2|{\cal A}_{\mathrm{B}}^{0-}|^2\times
\nonumber \\
&& \left [\frac{m_l^2}{(p_l\cdot q)^2}
        +\frac{M_{\pi}^2}{(p_2\cdot q)^2}-
        \frac{2p_2\cdot p_l}{(p_2\cdot q)(p_l\cdot q)}
        \right ]\,.
        \nonumber
\end{eqnarray}

The preceding expression is singular for vanishing momentum of the
soft photon. We shall attribute a small but non-vanishing mass to
the photon, $m_{\gamma}$, in order to regularize this singularity.

\subsection{The decay rate}
\label{sec:decay rate}

The $K_{\ell 4\gamma}$ differential decay rate is obtained by
squaring the matrix element (\ref{eq:bremsstrahlung amplitude}),
summing over spins and polarizations and integrating over the
following phase space,
\begin{eqnarray}
 d\Phi_{\gamma}
  &=& (2\pi )^4\delta^{(4)}(p_1+p_2+p_l+p_{\nu}+q-p)\times \nonumber \\
  & & \frac{d^3\boldsymbol{p}_1}{(2\pi )^32E_1}\,
        \frac{d^3\boldsymbol{p}_2}{(2\pi )^32E_2}\,
        \frac{d^3\boldsymbol{p}_l}{(2\pi )^32E_l}\,
        \frac{d^3\boldsymbol{p}_{\nu}}{(2\pi
        )^32|\boldsymbol{p}_{\nu}|}\,\frac{d^3\boldsymbol{q}}{(2\pi )^32|\boldsymbol{q}|}\,.
        \nonumber
\end{eqnarray}

Using the definition of bremsstrahlung integrals as given in the
appendix, the $K_{\ell 4\gamma}$ differential decay rate takes the following form in the soft photon approximation,
\begin{eqnarray}
d\Gamma_{\gamma}
  &=& -\dfrac{e^2}{2}\,\dfrac{1}{2M_K}\,d\Phi\sum_{\mathrm{spins}}|{\cal A}_{\mathrm{B}}^{0-}|^2\times
\nonumber \\ 
&& \left[ I(p_2,p_2,m_{\gamma},\omega )+I(p_l,p_l,m_{\gamma},\omega )-2I(p_2,p_l,m_{\gamma},\omega )\right] \,, \label{eq:kl4gddr}
\end{eqnarray}
where non singular $m_{\gamma}$ terms have been dropped out. 

\subsection{Cancellation of infrared divergencies}
\label{sec:Cancellation of infrared divergencies}

The infrared divergent part of $K_{\ell 4\gamma}$ differential
decay rate can be extracted from (\ref{eq:kl4gddr}) using the
definition of bremsstrahlung integrals from the appendix,
\begin{eqnarray}
d\Gamma_{\gamma}^{\mathrm{IR}}
&=& \frac{e^2}{4\pi^2}\,
\dfrac{1}{2M_K}\,d\Phi\sum_{\mathrm{spins}}|{\cal A}_{\mathrm{B}}^{0-}|^2\times
\nonumber \\ 
&& \left[ 1+p_l\cdot p_2\tau (-p_l,p_2,m_l,M_{\pi})\right] \ln m_{\gamma}^2\,. 
\end{eqnarray}
On the other hand, the infrared divergence coming from virtual
photon corrections to $F$ and $G$ form factors only can be read
off from Tab.~\ref{tab:1} and is denoted by $d\Gamma^{\mathrm{IR}}$. It is easy then to check that infrared divergencies cancel at the
level of differential decay rates,
$$
d\Gamma^{\mathrm{IR}}+d\Gamma_{\gamma}^{\mathrm{IR}}\,=\,0\,.
$$

\section{Perspectives}
\label{sec:perspectives} 

In this work we studied the decay process, $K^0\rightarrow\pi^0\pi^-\ell^+\nu_\ell$, taking into account Isospin breaking effects. These come mainly from electroweak interactions and generate corrections proportional to the fine structure constant, $\alpha$, and to the difference between up and down quark masses, $m_u-m_d$. 

The interest in this decay comes from the fact that the partial wave expansion of the corresponding form factors involves $\pi\pi$ scattering phase shifts. The latter can be related in a model-independent way to the $\pi\pi$ scattering lengths which are sensitive to the value of the quark condensate. Thus, a precise measurement of form factors should allow accurate determination of scattering lengths and, consequently, give precious information about the QCD vacuum structure. 

Scattering lengths are strong interaction quantities. On the other hand, any $K_{\ell 4}$ decay measurement contains contributions from all possible interactions, in particular, from electroweak ones. Therefore, it is primordial to have under control Isospin breaking effects in order to disentangle the strong interaction contribution from the measured form factors. The present work was guided by this motivation and, to this end, analytic expressions for form factors were obtained including Isospin breaking effects. These expressions are ultraviolet finite, scale independent, but infrared divergent. We showed that this divergence cancels out at the differential decay rate level if we take into account real soft photon emission. 

Our work should be completed by, 
\begin{itemize}
\item a parametrization of form factors in the presence of Isospin breaking,
\item a full treatment of the radiative decay, $K_{\ell 4\gamma}$.  
\end{itemize}

\begin{flushleft}
\textit{\textbf{Acknowledgements}}
\end{flushleft} 
I am grateful to Mark Knecht for pointing out to me the relevance of the subject and for his constant support. 

\appendix

\section{Loop integrals}
\label{app:loop_integrals}

We list here analytical expressions for scalar one-loop integrals and reduction of tensor integrals to scalar ones~\cite{Denner:1993kt}. Let $D$ be space-time dimension and $\eta^{\mu\nu}$ the metric tensor. For large momenta, loop integrals are divergent when $D=4$. A regularization should then be applied to treat this \textit{ultraviolet divergence}. The \textit{dimensional regularization} consists on calculating integrals for arbitrary $D$. Physical situations are recovered in the limit $D\rightarrow 4$. Infinities show up as poles in inverse powers of $\varepsilon\doteq 4-D>0$. These poles have to be absorbed (subtracted) by \textit{renormalization constants} in order to obtain finite (observable) results. This subtraction is not unique and it is a matter of taste to chose one of the different schemes. In the $\overline{\mathrm{MS}}$ scheme, one subtracts the following,
\begin{equation}
\overline{\lambda}\,\doteq\,-\dfrac{1}{32\pi^2}\left[ \dfrac{2}{\varepsilon}
+1-\gamma_{\mathrm{E}}+\ln (4\pi )\right] \,,
\end{equation}   
where $\gamma_{\mathrm{E}}$ is the \textit{Euler constant}. Another feature of dimensional regularization is the scale $\mu$. It has the dimension of a mass and maintains correct units while dimensionally regularizing integrals,
\begin{equation}
\int\dfrac{d^4l}{(2\pi )^4} \longrightarrow \mu^{4-D}\int\dfrac{d^Dl}{(2\pi )^D}\,.
\end{equation}  
Note that the scale $\mu$ is also absorbed by renormalization constants so that observables remain scale independent. 

Some integrals are divergent for vanishing momenta or masses. This kind of divergence is called \textit{infrared divergence} and can be treated in dimensional regularization quite as the ultraviolet one ($\varepsilon$ being negative herein). For the situation we are considering, infrared divergence is due to the presence of virtual photons. One can then think about a \textit{cutoff regularization} and assigns to the photon a fictitious infinitely small mass, $m_\gamma$, which play the role of the cutoff. Obviously, observables do not depend on $m_\gamma$. In fact, the infrared divergence due to virtual photons is cancelled by the one coming from real soft photon emission leading to cutoff independent quantities.     

In order to evaluate numerically the finite part of loop integrals scalar ones are expressed in terms of elementary functions like logarithms. The arguments of the latter can always be cast in the following compact form,
\begin{equation} \label{eq:argument} 
\sigma (z,m,m')\,\doteq\,\dfrac{1-\sqrt{1-\dfrac{4mm'}{z-(m-m')^2}}}
{1+\sqrt{1-\dfrac{4mm'}{z-(m-m')^2}}}\,,
\end{equation}
with $z$ a complex quantity whereas $m$ and $m'$ are real. For complex arguments, the logarithm is analytic and presents a cut structure given on the first Riemann sheet by,
\begin{equation}
\ln (x+i\varsigma )\,=\,\ln \vert x\vert +i\pi\Theta (-x)\mathrm{sgn}(\varsigma )\,,
\end{equation}
for real $x$ and infinitesimal $\varsigma$. Herein, $\Theta$ is the Heaviside function,
\begin{equation}
\Theta (x)\,\doteq\,\left\lbrace \begin{array}{ccc}
                                 1 & \mathrm{for} & x>0 \\
                                 0 & \mathrm{for} & x\leq 0 
                                 \end{array}\right.\,,
\end{equation} 
and sgn is the sign function,
\begin{equation}
\mathrm{sgn}(\varsigma )\,\doteq\,\left\lbrace \begin{array}{ccc}
                                              1 & \mathrm{for} & x>0 \\
                                             -1 & \mathrm{for} & x<0 
                                              \end{array}\right.\,.
\end{equation}  
For three-point and higher functions, more complicated (but well known) functions have to be used. This is the case, say, of dilogarithm,
\begin{equation}
\mathrm{Li}_2(z)\,\doteq\,-\int_0^1dt\,t^{-1}\,\ln (1-zt)\,.
\end{equation} 
The dilogarithm is analytic in $z$. That is, it develops an imaginary part for values of $z$ fixed by the cut structure of $\ln (1-z)$ as seen from the definition.  
From the foregoing one can easily prove the following identity for $x$ real such that $x\geq 1$, 
\begin{equation}
\mathrm{Li}_2(x+i\varsigma )-\mathrm{Li}_2(x-i\varsigma ) 
\,=\,2i\pi\,\mathrm{sgn}(\varsigma )\ln x\,.
\end{equation} 
For the convenience of giving compact expressions of loop integrals in the various regions of the complex momentum space we will introduce the logical function,
\begin{equation}
\mathrm{If}(\mathit{argument})
\,\doteq\,\left\lbrace \begin{array}{cc}
                       1 & \mathrm{if}~\mathit{argument}~\mathrm{is~true} \\
                       0 & \mathrm{if}~\mathit{argument}~\mathrm{is~false}
                       \end{array}\right.\,. 
\end{equation}  
The analytic structure of logarithms and dilogarithms is proving relevant for the determination of loop functions in physical regions. Let $p_i$ and $p_j$ be external momenta in a given amplitude and define the \textit{exchange energy} as, 
\begin{equation}
p_{ij}^2\,\doteq\,(p_i-p_j)^2\,,  \qquad i\,=\,1,\,2,\,3\,.
\end{equation} 
If $m_i$ and $m_j$ are internal masses then loop integrals are singular for those values given by Landau's equations,
\begin{equation}
p_{ij}^2\,=\,(m_i\mp m_j)^2\,,
\end{equation} 
corresponding respectively to \textit{pseudo} and \textit{normal thresholds} of the amplitude. Thresholds divide the momentum space into three regions,
$$
p_{ij}^2<(m_i-m_j)^2\,,\, (m_i-m_j)^2<p_{ij}^2<(m_i+m_j)^2\,,\, p_{ij}^2>(m_i+m_j)^2\,,
$$
which can be reached from each other by \textit{analytic continuation} from real to complex momentum values,
\begin{equation}
p_{ij}^2\longrightarrow p_{ij}^2+i\epsilon\,,
\end{equation}
with $\epsilon$ an infinitesimal positive quantity. To see how it works, let us consider the analytic continuation of (\ref{eq:argument}),
\begin{equation}
\sigma_{ij}\,\doteq\,\sigma\left( p_{ij}^2+i\epsilon ,m_i,m_j\right) \,.
\end{equation}
It is easy to show that the logarithm of the latter function takes the following form,
\begin{eqnarray}
\ln\left( \sigma_{ij}\right) 
&=& -\,\mathrm{If}\left( p_{ij}^2<(m_i-m_j)^2\right)\,\times
\nonumber \\ 
& & \ln\,\dfrac{\sqrt{(m_i+m_j)^2-p_{ij}^2}+\sqrt{(m_i-m_j)^2-p_{ij}^2}}
{\sqrt{(m_i+m_j)^2-p_{ij}^2}-\sqrt{(m_i-m_j)^2-p_{ij}^2}}
\nonumber \\ 
&+& 2i\,\mathrm{If}\left( (m_i-m_j)^2<p_{ij}^2<(m_i+m_j)^2\right) \,\times 
\nonumber \\ 
& & \mathrm{arctan}\,\dfrac{\sqrt{p_{ij}^2-(m_i-m_j)^2}}{\sqrt{(m_i+m_j)^2-p_{ij}^2}} 
\nonumber \\
&-& \mathrm{If}\left( p_{ij}^2>(m_i+m_j)^2\right) \,\times 
\nonumber \\ 
& & \left[ \ln\,\dfrac{\sqrt{p_{ij}^2-(m_i-m_j)^2}+\sqrt{p_{ij}^2-(m_i+m_j)^2}}
{\sqrt{p_{ij}^2-(m_i-m_j)^2}-\sqrt{p_{ij}^2-(m_i+m_j)^2}}-i\pi\right]\,. 
\label{eq:logarithm} 
\end{eqnarray}  
Finally, the following notations are usefull for the reduction of vector and tensor integrals to scalar ones~\cite{Denner:2002ii} ,
\begin{equation}
N_0\,\doteq\,l^2-m_0^2+i\epsilon\,, 
\qquad N_i\,\doteq\,(p_i+l)^2-m_i^2+i\epsilon\,,
\end{equation} 
\begin{equation}
2p_i\cdot l\,=\,N_i-N_0-f_i\,, \qquad f_i\,\doteq\,p_i^2-m_i^2+m_0^2\,.
\end{equation}

\subsection{One-point functions}

The one-point functions or tadpole integrals are defined by,
\begin{equation}
\left\lbrace A_0, A^\mu ,A^{\mu\nu}\right\rbrace (m_0)
\,\doteq\,
-i\mu^{4- D}\int\dfrac{d^Dl}{(2\pi )^D}\,N_0^{-1}\left\lbrace 1, l^\mu , l^\mu l^\nu\right\rbrace \,.
\end{equation} 
In dimensional regularization the scalar integral reads,
\begin{equation}
A_0\,=\,m_0^2\left[ -2\overline{\lambda}
-\dfrac{1}{16\pi^2}\,\ln\left( \dfrac{m_0^2}{\mu^2}\right) \right] \,.
\end{equation}
The vector and tensor one-point functions are found to be,
\begin{equation}
A^\mu\,=\,0\,, \qquad 
A^{\mu\nu}\,=\,\dfrac{m_0^2}{8}\left[ 2A_0(m_0)+\dfrac{m_0^2}{16\pi^2}\right] \,.
\end{equation} 

\subsection{Two-point functions}

The two-point functions are defined by,
\begin{eqnarray}
&& \left\lbrace B_0,B^\mu ,B^{\mu\nu}\right\rbrace (p_1,m_0,m_1)\,\doteq \, 
\nonumber \\ 
&& \qquad\qquad -i\mu^{4-D}\int\dfrac{d^Dl}{(2\pi )^D}\,(N_0\,N_1)^{-1}
                \left\lbrace 1, l^\mu , l^\mu l^\nu\right\rbrace \,.
\end{eqnarray}
The scalar integral reads~\cite{Bohm:1986rj} ,
\begin{eqnarray}
B_0 
&=& \dfrac{1}{2}\,\dfrac{A_0(m_0)}{m_0^2}+\dfrac{1}{2}\,\dfrac{A_0(m_1)}{m_1^2} \nonumber \\
&+& \dfrac{1}{16\pi^2}\left\lbrace 1-\dfrac{m_0^2-m_1^2}{p_1^2}\,\ln\dfrac{m_0}{m_1}\right. 
\nonumber \\ 
&+& \dfrac{1}{p_1^2}\,\sqrt{(m_0+m_1)^2-p_1^2}\sqrt{(m_0-m_1)^2-p_1^2}\,\times 
\nonumber \\ 
& & \ln\,\dfrac{\sqrt{(m_0+m_1)^2-p_1^2}+\sqrt{(m_0-m_1)^2-p_1^2}}
{\sqrt{(m_0+m_1)^2-p_1^2}-\sqrt{(m_0-m_1)^2-p_1^2}}\,\times 
\nonumber \\ 
& & \mathrm{If}\left( p_1^2<(m_0-m_1)^2\right) 
\nonumber \\ 
&-& \dfrac{2}{p_1^2}\,\sqrt{(m_0+m_1)^2-p_1^2}\,\sqrt{p_1^2-(m_0-m_1)^2}\,\times 
\nonumber \\ 
& & \mathrm{arctan}\,\dfrac{\sqrt{p_1^2-(m_0-m_1)^2}}{\sqrt{(m_0+m_1)^2-p_1^2}}\,\times 
\nonumber \\ 
& & \mathrm{If}\left( (m_0-m_1)^2<p_1^2<(m_0+m_1)^2\right) 
\nonumber \\ 
&-& \dfrac{1}{p_1^2}\,\sqrt{p_1^2-(m_0+m_1)^2}\,\sqrt{p_1^2-(m_0-m_1)^2}\,\times 
\nonumber \\ 
& & \left[ \ln\,\dfrac{\sqrt{p_1^2-(m_0-m_1)^2}+\sqrt{p_1^2-(m_0+m_1)^2}}
{\sqrt{p_1^2-(m_0-m_1)^2}-\sqrt{p_1^2-(m_0+m_1)^2}}-i\pi\right]\,\times 
\nonumber \\ 
& & \mathrm{If}\left( p_1^2>(m_0+m_1)^2\right)\bigg\}\,.
\end{eqnarray}  
The vector two-point function is written as,
\begin{equation}
B^\mu\,\doteq\,p_1^\mu\,B_1\,,
\end{equation} 
with the coefficient,
\begin{equation}
2p_1^2B_1\,=\,A_0(m_0)-A_0(m_1)-f_1B_0\,.
\end{equation} 
The tensor two-point function possesses the following decomposition,
\begin{equation}
B^{\mu\nu}\,\doteq\,\eta^{\mu\nu}B_{00}+p_1^\mu p_1^\nu B_{11}\,,
\end{equation} 
with the coefficients,
\begin{eqnarray}
18B_{00}
&=& 3A_0(m_1)
\nonumber \\ 
&+& 6m_0^2B_0+3f_1B_1+\dfrac{1}{16\pi^2}\left( 4m_0^2+2m_1^2-f_1\right) \,, \\ 
18p_1^2B_{11}
&=& 6A_0(m_1)
\nonumber \\ 
&-& 6m_0^2B_0-12f_1B_1-\dfrac{1}{16\pi^2}\left( 4m_0^2+2m_1^2-f_1\right) \,.
\end{eqnarray} 
Two-point functions are ultraviolet divergent. It is convenient to define ultraviolet finite parts by subtracting the corresponding poles. The finite parts are noted with the superscript "$r$" like renormalized and are defined as follows,
\begin{equation}
B_0\,\doteq\,B_0^r-2\overline{\lambda}\,, \quad B_1\,\doteq\,B_1^r-\overline{\lambda}\,, 
\end{equation}
\begin{equation} B_{00}\,\doteq\,B_{00}^r+\dfrac{1}{6}\,(p_1^2-3m_0^2-3m_1^2)\,\overline{\lambda}\,, \quad B_{11}\,\doteq\,B_{11}^r-\dfrac{2}{3}\,\overline{\lambda}\,.
\end{equation} 

\subsection{Three-point functions}

The three-point functions are defined by,
\begin{eqnarray}
&& \left\lbrace C_0,C^\mu ,C^{\mu\nu}\right\rbrace (p_1,p_2,m_0,m_1,m_2)\,\doteq\, 
\nonumber \\ 
&& \qquad\qquad\qquad -i\mu^{4-D}\int\dfrac{d^Dl}{(2\pi )^D}\,(N_0\,N_1\,N_2)^{-1}
\left\lbrace 1, l^\mu , l^\mu l^\nu\right\rbrace \,.
\end{eqnarray}
The general expression for the scalar integral~\cite{'tHooft:1979xw} is complicated especially for separating real and imaginary parts. In order to make this separation more transparent we will give a one-dimensional integral representation of the scalar integral~\cite{Binoth:2002xh},
\begin{eqnarray}
C_0
&=& -\dfrac{1}{16\pi^2}\int_0^1dx\left[ \mathcal{F}(x;p_1^2,p_{12}^2,p_2^2,m_1^2,m_2^2,m_0^2)\right. 
\nonumber \\ 
&& \qquad\qquad\qquad +\mathcal{F}(x;p_2^2,p_1^2,p_{12}^2,m_0^2,m_1^2,m_2^2)
\nonumber \\
&& \qquad\qquad\qquad \left.+\mathcal{F}(x;p_{12}^2,p_2^2,p_1^2,m_2^2,m_0^2,m_1^2)\right] \,.
\end{eqnarray} 
Herein,
\begin{eqnarray}
&& \dfrac{d^2-\Delta}{4a}\,\mathcal{F}(x;p_1^2,p_{12}^2,p_2^2,m_1^2,m_2^2,m_0^2)\,\doteq \nonumber \\ 
&& \ln (2a+b+d)-\ln (b+d) 
\nonumber \\ 
&+& \mathrm{If}(\Delta <0)\left\lbrace \dfrac{1}{2}\,\ln c-\dfrac{1}{2}\,\ln (a+b+c)\right. 
\nonumber \\ 
&+& \dfrac{d}{\sqrt{-\Delta}}\left[ \mathrm{arctan}\left( \dfrac{\sqrt{-\Delta}}{b}\right) \right. 
\nonumber \\ 
&-& \left.\left.\mathrm{arctan}\left( \dfrac{\sqrt{-\Delta}}{2a+b}\right) +\pi\,\mathrm{If}(b<0<2a+b)\right] \right\rbrace 
\nonumber \\ 
&+& \mathrm{If}(\Delta>0)\left\lbrace \dfrac{d-\sqrt{\Delta}}{2\sqrt{\Delta}}\left[ \ln\vert 2a+b-\sqrt{\Delta}\vert\right.\right. 
\nonumber \\ 
&-& \left.\ln\vert b-\sqrt{\Delta}\vert -i\pi\,\mathrm{If}(b<\sqrt{\Delta}<2a+b)\right] 
\nonumber \\ 
&& \qquad -\dfrac{d+\sqrt{\Delta}}{2\sqrt{\Delta}}\left[ \ln\vert 2a+b+\sqrt{\Delta}\vert\right. 
\nonumber \\ 
&-& \left.\left.\ln\vert b+\sqrt{\Delta}\vert +i\pi\,\mathrm{If}(b<-\sqrt{\Delta}<2a+b)\right] \right\rbrace \,,  
\end{eqnarray}
with,
\begin{eqnarray}
a
&=& m_2^2\,, 
\\
b
&=& (m_1^2+m_2^2-p_{12}^2)x+m_2^2+m_0^2-p_2^2\,, 
\\
c
&=& m_1^2x^2+(m_1^2+m_0^2-p_1^2)x+m_0^2\,, 
\\ 
d
&=& 2a(1+x)-b\,, 
\\ 
\Delta
&=& b^2-4ac+i\epsilon\,.
\end{eqnarray}   
Obviously, this representation is appropriate for direct numerical integration. The vector three-point function is written as,
\begin{equation}
C^\mu\,\doteq\,p_1^\mu\,C_1+p_2^\mu\,C_2\,.
\end{equation} 
The coefficients are given in a compact form,
\begin{equation}
C_i\,=\,\sum_{n=1}^2\left( \mathcal{Z}_2\right) _{in}^{-1}\mathcal{C}_n\,, \qquad 
i\,=\,1,\,2\,,
\end{equation}
where,
\begin{equation}
\mathcal{Z}_2\,\doteq\,\left( \begin{array}{cc}
                              2p_1^2 & 2p_1\cdot p_2 \\
                              2p_1\cdot p_2 & 2p_2^2
                              \end{array}\right) \,,
\end{equation}
and,
\begin{eqnarray}
\mathcal{C}_1
&=& B_0(p_2,m_0,m_2)-B_0(p_2-p_1,m_1,m_2)-f_1C_0\,, 
\\ 
\mathcal{C}_2
&=& B_0(p_1,m_0,m_1)-B_0(p_2-p_1,m_1,m_2)-f_2C_0\,.
\end{eqnarray}     
The tensor three-point function possesses the following decomposition,
\begin{equation}
C^{\mu\nu}\,\doteq\,\eta^{\mu\nu}C_{00}+\sum_{j,k=1}^2p_j^{\mu}p_k^{\nu}C_{jk}\,.
\end{equation} 
The coefficients are found to be,
\begin{eqnarray}
4C_{00}
&=& B_0(p_2-p_1,m_1,m_2)+2m_0^2C_0+f_1C_1+f_2C_2+\dfrac{1}{16\pi^2}\,, 
\\ 
C_{ij}
&=& \sum_{n=1}^2\left( \mathcal{Z}_2\right) _{jn}^{-1}(\mathcal{C}_{in}-2\delta_{in}C_{00})\,, \qquad i,\,j\,=\,1,\,2\,,
\end{eqnarray}
where,
\begin{eqnarray}
\mathcal{C}_{11}
&=& B_0(p_2-p_1,m_1,m_2)+B_1(p_2-p_1,m_1,m_2)-f_1C_1\,, 
\\ 
\mathcal{C}_{12}
&=& B_1(p_1,m_0,m_1)-B_1(p_1-p_2,m_2,m_1)-f_2C_1\,, 
\\
\mathcal{C}_{21}
&=& B_1(p_2,m_0,m_2)-B_1(p_2-p_1,m_1,m_2)-f_1C_2\,, 
\\
\mathcal{C}_{22}
&=& B_0(p_1-p_2,m_2,m_1)+B_1(p_1-p_2,m_2,m_1)-f_2C_2\,.
\end{eqnarray} 
Only the tensor coefficient $C_{00}$ is ultraviolet divergent. The finite part is defined by the following subtraction,
\begin{equation}
C_{00}\,\doteq\,C_{00}^r-\dfrac{1}{2}\,\overline{\lambda}\,.
\end{equation}  

\subsection{Four-point functions}

The four-point functions are defined by,
\begin{eqnarray}
&& \left\lbrace D_0,D^\mu ,D^{\mu\nu}\right\rbrace (p_1,p_2,p_3,m_0,m_1,m_2,m_3)\,\doteq\, 
\nonumber \\ 
&& \qquad\qquad\qquad -i\mu^{4-D}\int\dfrac{d^Dl}{(2\pi )^D}\,(N_0\,N_1\,N_2\,N_3)^{-1}
\left\lbrace 1, l^\mu , l^\mu l^\nu\right\rbrace \,.
\end{eqnarray}
The expression for the scalar integral is known but difficult to handle~\cite{Denner:1991qq}. Since we are interested only in the case, $m_0\,=\,m_{\gamma}$, $p_1^2\,=\,m_1^2$, $p_2^2\,=\,m_2^2$, the general expression will not be reproduced here. The vector four-point function is written as,
\begin{equation}
D^\mu\,\doteq\,p_1^\mu\,D_1+p_2^\mu\,D_2+p_3^\mu\,D_3\,.
\end{equation} 
The coefficients are given in a compact form,
\begin{equation}
D_i\,=\,\sum_{n=1}^3\left( \mathcal{Z}_3\right) _{in}^{-1}\mathcal{D}_n\,, \qquad 
i\,=\,1,\,2\,3\,,
\end{equation}
where,
\begin{equation}
\mathcal{Z}_3\,\doteq\,\left( \begin{array}{ccc}
                              2p_1^2 & 2p_1\cdot p_2 & 2p_1\cdot p_3 \\
                              2p_1\cdot p_2 & 2p_2^2 & 2p_2\cdot p_3 \\ 
                              2p_1\cdot p_3 & 2p_2\cdot p_3 & 2p_3^2
                              \end{array}\right) \,,
\end{equation}
and,
\begin{eqnarray}
\mathcal{D}_1
&=& C_0(p_2,p_3,m_0,m_2,m_3)
\nonumber \\ 
&-& C_0(p_2-p_1,p_3-p_1,m_1,m_2,m_3)-f_1D_0\,, 
\\ 
\mathcal{D}_2
&=& C_0(p_1,p_3,m_0,m_1,m_3)
\nonumber \\ 
&-& C_0(p_2-p_1,p_3-p_1,m_1,m_2,m_3)-f_2D_0\,, 
\\
\mathcal{D}_3
&=& C_0(p_1,p_2,m_0,m_1,m_2)
\nonumber \\ 
&-& C_0(p_2-p_1,p_3-p_1,m_1,m_2,m_3)-f_3D_0\,.
\end{eqnarray}     
The tensor four-point function possesses the following decomposition,
\begin{equation}
D^{\mu\nu}\,\doteq\,\eta^{\mu\nu}D_{00}+\sum_{j,k=1}^3p_j^{\mu}p_k^{\nu}D_{jk}\,.
\end{equation} 
The coefficients are found to be,
\begin{eqnarray}
2D_{00}
&=& C_0(p_2-p_1,p_3-p_1,m_1,m_2,m_3)+2m_0^2D_0+\sum_{n=1}^3f_nD_n\,, 
\\ 
D_{ij}
&=& \sum_{n=1}^3\left( \mathcal{Z}_3\right) _{jn}^{-1}(\mathcal{D}_{in}-2\delta_{in}D_{00})\,, \qquad i,\,j\,=\,1,\,2\,3\,,
\end{eqnarray}
where,
\begin{eqnarray}
\mathcal{D}_{11}
&=& C_0(p_2-p_1,p_3-p_1,m_1,m_2,m_3)+C_1(p_2-p_1,p_3-p_1,m_1,m_2,m_3)
\nonumber \\ 
&+& C_2(p_2-p_1,p_3-p_1,m_1,m_2,m_3)-f_1D_1\,, 
\\ 
\mathcal{D}_{12}
&=& C_1(p_1,p_3,m_0,m_1,m_3)
\nonumber \\ 
&-& C_2(p_3-p_2,p_1-p_2,m_2,m_3,m_1)-f_2D_1\,, 
\\
\mathcal{D}_{13}
&=& C_1(p_1,p_2,m_0,m_1,m_2)
\nonumber \\ 
&-& C_1(p_1-p_3,p_2-p_3,m_3,m_1,m_2)-f_3D_1\,, 
\\
\mathcal{D}_{21}
&=& C_1(p_2,p_3,m_0,m_2,m_3)
\nonumber \\ 
&-& C_1(p_2-p_1,p_3-p_1,m_1,m_2,m_3)-f_1D_2\,, 
\\
\mathcal{D}_{22}
&=& C_0(p_3-p_2,p_1-p_2,m_2,m_3,m_1)+C_1(p_3-p_2,p_1-p_2,m_2,m_3,m_1)
\nonumber \\ 
&+& C_2(p_3-p_2,p_1-p_2,m_2,m_3,m_1)-f_2D_2\,, 
\\ 
\mathcal{D}_{23}
&=& C_2(p_1,p_2,m_0,m_1,m_2)
\nonumber \\ 
&-& C_2(p_1-p_3,p_2-p_3,m_3,m_1,m_2)-f_3D_2\,, 
\\  
\mathcal{D}_{31}
&=& C_2(p_2,p_3,m_0,m_2,m_3)
\nonumber \\ 
&-& C_2(p_2-p_1,p_3-p_1,m_1,m_2,m_3)-f_1D_3\,, 
\\   
\mathcal{D}_{32}
&=& C_2(p_1,p_3,m_0,m_1,m_3)
\nonumber \\ 
&-& C_1(p_3-p_2,p_1-p_2,m_2,m_3,m_1)-f_2D_3\,, 
\\
\mathcal{D}_{33}
&=& C_0(p_1-p_3,p_2-p_3,m_3,m_1,m_2)+C_1(p_1-p_3,p_2-p_3,m_3,m_1,m_2)
\nonumber \\ 
&+& C_2(p_1-p_3,p_2-p_3,m_3,m_1,m_2)-f_3D_3\,.
\end{eqnarray} 

\subsection{Infrared divergent integrals}

From all integrals cited before only the scalar three- and four-point integrals are infrared divergent for the particular case,
\begin{equation}
m_0\,=\,m_{\gamma}\,, \qquad p_1^2\,=\,m_1^2\,, \qquad p_2^2\,=\,m_2^2\,. 
\end{equation} 
Furthermore, these integrals are related by,
\begin{equation}
C_0(p_1,p_2,m_{\gamma},m_1,m_2)\,=\,\lim_{m_3^2\rightarrow\infty}\left[ -m_3^2D_0(p_1,p_2,p_3,m_{\gamma},m_1,m_2,m_3)\right] \,.
\end{equation} 
In this subsection, the integrals in question will be denoted $C_0$ and $D_0$, respectively, and read~\cite{Beenakker:1990jr} ,
\begin{eqnarray}
C_0
&=& \dfrac{1}{16\pi^2}\,\dfrac{1}{m_1m_2}\,
\dfrac{\sigma_{12}}{1-\sigma_{12}^2}\times
\nonumber \\ 
&& \left\lbrace \ln\left( \sigma_{12}\right) \left[ 2\ln\left( 1-\sigma_{12}^2\right) -\dfrac{1}{2}\,\ln\left( \sigma_{12}\right) -\ln\left( \dfrac{m_{\gamma}^2}{m_1m_2}\right)\right] \right. 
\nonumber \\ 
&& -\dfrac{\pi^2}{6}+\mathrm{Li}_2\left( \sigma_{12}^2\right) +\dfrac{1}{2}\,\ln^2\left( \dfrac{m_1}{m_2}\right) 
\nonumber \\ 
&& \left.+\mathrm{Li}_2\left( 1-\sigma_{12}\,\dfrac{m_1}{m_2}\right) +\mathrm{Li}_2\left( 1-\sigma_{12}\,\dfrac{m_2}{m_1}\right)\right\rbrace \,,
\\ 
D_0
&=& \dfrac{1}{16\pi^2}\,\dfrac{1}{m_1m_2}\,\dfrac{1}{p_3^2-m_3^2}\,
\dfrac{\sigma_{12}}{1-\sigma_{12}^2}\times
\nonumber \\ 
&& \left\lbrace 2\ln\left( \sigma_{12}\right) \left[ \ln\left( 1-\sigma_{12}^2\right) -\ln\left( \dfrac{m_3m_{\gamma}}{m_3^2-p_3^2-i\epsilon}\right) \right] \right. 
\nonumber \\ 
&& +\dfrac{\pi^2}{2}+\mathrm{Li}_2\left( \sigma_{12}^2\right) +\ln^2\left( \sigma_{13}\right) +\ln^2\left( \sigma_{23}\right) -\sum_{j,k=-1,1}\left[ \mathrm{Li}_2\left( \sigma_{12}\sigma_{13}^k\sigma_{23}^j\right) \right. 
\nonumber \\ 
&& \left.\left.+\left( \ln\left( \sigma_{12}\right) +\ln\left( \sigma_{13}^k\right) +\ln\left( \sigma_{23}^j\right) \right) \ln\left( 1-\sigma_{12}\sigma_{13}^k\sigma_{23}^j\right) \right] \right\rbrace \,.
\end{eqnarray} 
The $\ln m_{\gamma}$ terms in the preceding formulae generate infrared divergence in the virtual photon correction to the amplitude. This divergence is cancelled by the one coming from the associated soft real photon emission originating from the bremsstrahlung integrals,
\begin{equation}
I(p_1,p_2,m_{\gamma},\omega )\,\doteq\,\int_{m_{\gamma}}^{\omega}\dfrac{d^3{\boldsymbol q}}{(2\pi )^32|{\boldsymbol q}|}\,\dfrac{2p_1\cdot p_2}{(p_1\cdot q)(p_2\cdot q)}\,.
\end{equation} 
The expression of bremsstrahlung integrals in terms of logarithms and dilogarithms reads,
\begin{eqnarray}
I
&=& \dfrac{1}{4\pi^2}\,\dfrac{\alpha p_1\cdot p_2}{\alpha p_1\cdot p_2-p_2^2}\left\lbrace \ln\dfrac{2\alpha p_1\cdot p_2-p_2^2}{p_2^2}\,\ln\dfrac{2\omega}{m_{\gamma}}\right. 
\nonumber \\ 
&+& \dfrac{1}{4}\,\ln^2\dfrac{p_1^0-|{\boldsymbol p}_1}{p_1^0+|{\boldsymbol p}_1}-\dfrac{1}{4}\,\ln^2\dfrac{p_2^0-|{\boldsymbol p}_2}{p_2^0+|{\boldsymbol p}_2}
\nonumber \\ 
&+& \mathrm{Li}_2\left( 1-\dfrac{\alpha}{\beta}\left( p_1^0+|{\boldsymbol p}_1|\right) \right) -\mathrm{Li}_2\left( 1-\dfrac{1}{\beta}\left( p_2^0+|{\boldsymbol p}_2|\right) \right)
\nonumber \\ 
&+& \left.\mathrm{Li}_2\left( 1-\dfrac{\alpha}{\beta}\left( p_1^0-|{\boldsymbol p}_1|\right) \right) -\mathrm{Li}_2\left( 1-\dfrac{1}{\beta}\left( p_2^0-|{\boldsymbol p}_2|\right) \right)\right\rbrace \,,
\end{eqnarray} 
where,
\begin{equation}
\beta\,\doteq\,\dfrac{\alpha p_1\cdot p_2-p_2^2}{\alpha p_1^0-p_2^0}\,,
\end{equation} 
and $\alpha$ is defined through,
\begin{equation}
\alpha^2p_1^2-2\alpha p_1\cdot p_2+p_2^2\,, \qquad \dfrac{\alpha p_1^0-p_2^0}{p_2^0}\,>\,0\,.
\end{equation}
For the particular case, $p_1=p_2$ the integral simplifies to,
\begin{equation}
I(p_1,p_1,m_{\gamma},\omega )\,=\,\dfrac{1}{4\pi^2}\left( \ln\dfrac{4\omega^2}{m_{\gamma}^2}+\dfrac{p_1^0}{|{\boldsymbol p}_1|}\,\ln\dfrac{p_1^0-|{\boldsymbol p}_1|}{p_1^0+|{\boldsymbol p}_1|}\right) \,.
\end{equation} 
In order to make explicit the cancellation of infrared divergence in the amplitude we will use a compact unified notation for the $\ln m_{\gamma}$ terms in the integrals $C_0$, $D_0$ and $I$. To this end, define the $\tau$ function to be,
\begin{equation}
\tau (p_1,p_2,m_1,m_2)\,\doteq\,\int_0^1
\dfrac{dx}{p_{12}^2x^2-(p_{12}^2+m_1^2-m_2^2)x+m_1^2}\,.
\end{equation}
The $\tau$ function can be expressed in terms of the $\sigma$ function as follows,
\begin{equation}
\tau (p_1,p_2,m_1,m_2)\,=\,-\dfrac{2}{m_1m_2}\,
\dfrac{\sigma_{12}}{1-\sigma_{12}^2}\,\ln\left( \sigma_{12}\right)\,,
\end{equation}
where the logarithm can be read from (\ref{eq:logarithm}). In terms of the $\tau$ function, the infrared divergence in the three-point function, four-point function, and bremsstrahlung integral reads,
\begin{eqnarray}
C_0
&\longrightarrow & \dfrac{1}{32\pi^2}\,\tau (p_1,p_2,m_1,m_2)\,\ln m_{\gamma}^2\,,
\\ 
D_0
&\longrightarrow & \dfrac{1}{32\pi^2}\,\dfrac{1}{p_3^2-m_3^2}\,\tau (p_1,p_2,m_1,m_2)\,\ln m_{\gamma}^2\,,
\\ 
I
&\longrightarrow & -\dfrac{1}{4\pi^2}\,p_1\cdot p_2\,\tau (p_1,p_2,m_1,m_2)\,\ln m_{\gamma}^2\,.
\end{eqnarray}

\bibliographystyle{unsrt}

\bibliography{paper}

\begin{table}[p]
\begin{center}
\begin{tabular}{cc}
\hline
\textbf{diagram} & $\delta G$ \\
\hline & \\
\textbf{3. (a)} & $-\displaystyle\frac{e^2}{8\pi^2}\,\ln m_{\gamma}^2$ \\ & \\
\textbf{4. (c)} & $-\displaystyle\frac{e^2}{8\pi^2}\,p_l\cdot p_2\tau
(-p_l,p_2,m_l,M_{\pi})\ln
        m_{\gamma}^2$ \\ & \\
\hline
\end{tabular}
\end{center}
\caption{\label{tab:1} Infrared divergent part of the corrected
$f$ and $g$ form factors due to virtual photon corrections. The
contribution from \textbf{diagram 3. (a)} comes from wave function
renormalization of external charged particles, $\pi^-$, and $\ell^+$.}
\end{table}

\end{document}